
\input harvmac.tex
\input epsf.tex


\def \lij{l_{ij}}
\def \siint{\int \prod_i {ds_i\over 2\pi}}
\def \riint{\int \prod_i {dr_i\over 2\pi}}
\def \symf{{1\over \prod_{i,j}\lij!\prod_i P_i!2^{P_i}}}
\def \nf{{1\over N!}}
\def \gpi{\left[ G(0)\right]^{\sum_i P_i}}
\def \glij{\prod_{i<j}\left[G(s_i-s_j)\right]^{\lij}}
\def \part#1{{\partial \over \partial{#1}}}
\def \dtp#1{{d{#1}\over 2\pi}}
\def \hs#1{h_{\lij,P_k}({#1})}
\def \hsa#1{\tilde h_{\lij,P_k}({#1},\alpha)}
\def \hsat#1{\tilde h_{\tilde \lij,\tilde P_k}({#1},\alpha)}
\def \hst#1#2{h_{\lij,P_k}({#1,#2})}
\def \hsta#1#2{\tilde h_{\lij,P_k}({#1,#2,\alpha})}
\def \hstat#1#2{\tilde h_{\tilde \lij,\tilde P_k}({#1,#2,\alpha})}
\def \de{\Delta_\epsilon}
\def \detwo{\Delta_{2\epsilon}}

\def \vp{\vec \phi}
\def \vk{\vec k}
\def \gmn{G^{\mu\nu}}
\def \gmm{G^{\mu\mu}}
\def \gmme{\gmm_{2\epsilon}}
\def \gxx{G_{xx}}
\def \gyy{G_{yy}}
\def \gxy{G_{xy}}
\def \gyx{G_{yx}}
\def \emn{\epsilon^{\mu\nu}}
\def \aab{{\alpha\over\alpha^2 + \beta^2}}
\def \bab{{\beta\over\alpha^2 + \beta^2}}

\def \der#1{{d\over d{#1}}}

\def\sign{{\rm sign}}

\def\apm{\alpha^{\prime}}

\def\e{\epsilon}

\def\exp{{\rm exp}}

\def\sinh{{\rm sinh}}
\def\cosh{{\rm cosh}}
\def \sign{{\rm sign}}

\def\vX{\vec X}

\def\epiqx0#1{e^{i \vec q_{#1} \cdot \vX_0}}

\def\sqr#1#2{{\vbox{\hrule height.#2pt \hbox{\vrule width.#2pt
    height#1pt \kern#1pt \vrule width.#2pt} \hrule height.#2pt}}}
\def\dal{\mathchoice\sqr64\sqr64\sqr{4.2}3\sqr33}
%
%
\lref\dissqm{A.~O.~Caldeira and A.~J.~Leggett, Physica  {\bf 121A}(1983) 587;
Phys. Rev. Lett. {\bf 46} (1981) 211; Ann. of Phys. {\bf 149} (1983) 374.}
\lref\cgcdef{C.~G.~Callan and D.~Freed, Nucl. Phys {\bf B374}(1992)543.}
\lref\larsthesis{L. Thorlacius, ``String Theory on the Edge'', Princeton
Ph.D. Thesis, (1989).}
\lref\osdqm{C.G. Callan, L. Thorlacius,  Nucl. Phys. {\bf B329} (1990) 117.}
\lref\cff{C.~G.~Callan, A.~G.~Felce and D.~E.~Freed, Nucl. Phys. {\bf B392}
(1993) 551}
\lref\CTrivi{C.~G.~Callan, L.~Thorlacius, Nucl.  Phys. {\bf B319} (1989) 133.}
\lref\clny{C.~G.~Callan, C.~Lovelace, C.~R.~Nappi, and S.~A.~Yost,
Nucl. Phys. {\bf B293} (1987) 83; Nucl. Phys. {\bf B308} (1988) 221.}
\lref\schmid{A.~Schmid, Phys. Rev. Lett. {\bf 51} (1983) 1506.}
\lref\ghm{F.~Guinea, V.~Hakim and A.~Muramatsu, Phys. Rev. Lett.
{\bf 54} (1985) 263.}
\lref\fisher{M.~P.~A.~Fisher and W.~Zwerger, Phys. Rev. {\bf B32} (1985) 6190.}
\lref\klebanov{I.~Klebanov and L.~Susskind, Phys. Lett. {\bf B200} (1988) 446.}
\lref\soda{J.~Soda,``Renormalization Group Flow and String
Dynamics'',{\it Proceedings of the Summer Workshop in Superstrings},
Tsukuba, Japan (1988) 132.}
\lref\Larsnote{L.~Thorlacius, private communication.}
\lref\das{S.~Das and B.~Sathiapalan, Phys. Rev. Lett. {\rm 56}(1986)
2664.}
\lref\bkopen{M.~Bershadsky and D.~Kutasov, Phys. Lett. {\bf B274}(1992)331.}
\lref\freed{D.~E.~Freed, "Contact Terms and Duality Symmetry in the Critical
Dissipative Hofstadter Model'', CTP\#2170, to appear in Nucl. Phys. {\bf B}.}
\lref\amit{D.~J.~Amit, Y.~Y.~Goldschmidt and G.~Grinstein,  J. Phys.
{\bf A13},(1980)585.}
\lref\stamp{Y.-C.~Chen and P.~C.~E.~Stamp, in preparation.}
\lref\charged{D.~E.~Freed, "Fermionization and the Free Energy of
Charged 1-D Gases with Logarithmic Interactions", in preparation.}
%
{\nopagenumbers

\baselineskip 12pt plus 1pt minus 1pt
\centerline{\bf REPARAMETRIZATION INVARIANCE}
\smallskip
\centerline{{\bf IN SOME NON-LOCAL 1-D FIELD THEORIES}\footnote{*}{This
work is supported in part by National Science Foundation grant \#87-8447
and funds provided by the U. S. Department of Energy (D.O.E.) under contract
\#DE-AC02-76ER03069.}}
\vskip 24pt
\centerline{Denise E.~Freed}
\vskip 12pt
\centerline{\it Center for Theoretical Physics}
\centerline{\it Laboratory for Nuclear Science}
\centerline{\it and Department of Physics}
\centerline{\it Massachusetts Institute of Technology}
\centerline{\it Cambridge, Massachusetts\ \ 02139\ \ \ U.S.A.}
\vskip .7in
\centerline{\bf Abstract}
\smallskip
In this paper we consider 1-D non-local field theories with a
particular $1/r^2$ interaction, a constant gauge field and an
arbitrary scalar potential.  We show that any such theory that is
at a renormalization group fixed point also satisfies an infinite
set of reparametrization invariance Ward identities.  We also
prove that, for special values of the gauge field, the value of
the potential that satisfies the Ward identities to first order
in the potential strength remains a solution to all orders in the
potential strength, summed over all loops.  These theories are of
interest because they describe dissipative quantum mechanics with
an arbitrary potential and a constant magnetic field.  They also
give solutions to open string theory in the presence of a uniform
gauge field and an arbitrary tachyon field.
\vskip .8in
\centerline{Submitted to: {\it Nuclear Physics B\/}}
\vfill
\vskip -12pt
\noindent CTP\#2241\hfill September 1993
\eject}
\pageno=1
\baselineskip 12pt plus 1pt minus 1pt

\newsec{Introduction}
In 2-D statistical systems, scale invariance at a second order or higher
phase transition and locality generally lead to conformal invariance.
One dimensional systems can also have phase transitions, but only if they
are non-local.  Thus conformal invariance is not expected to be a symmetry
of critical 1-D theories.  However, it is
possible that they could exhibit some other enhanced symmetry related to
reparametrizations of the theory.  One example of a non-local 1-D system
that undergoes phase transitions is dissipative quantum mechanics, as
described by the Caldeira-Leggett model \dissqm.  The functional
integral for this theory is also used to obtain the boundary state in
open string theory \osdqm.
This boundary state describes an open string where the field is free on the
interior of the 2-D world sheet and all the interactions take place at the
1-D boundary.

Callan and Thorlacius have shown \CTrivi\ that in order to be a solution
to open string theory, the boundary state must satisfy a set of Ward
identities that reflect the reparmetrization invariance of the string theory.
Even though the 1-D field theory is not manifestly reparametrization invariant,
in order to correspond to a boundary state it must possess a ``hidden"
reparametrization symmetry, given by the Ward identities.  The question
arises as to whether we can use this symmetry to classify and construct 1-D
critical dissipative theories, just as conformal symmetry was used to
classify 2-D critical theories.  Also, it is natural to ask whether, for 1-D
theories, scale invariance of the theory implies that the theory satisfies
the Ward identities.  We will address this latter question in this paper
for the case of an arbitrary tachyon potential and a uniform gauge field.
For the dissipative quantum system, this corresponds to considering an
arbitrary potential and a uniform magnetic field.

Callan and Thorlacius have shown to one loop that for an arbitrary gauge
field the fixed point of the renormalization group also satisfies the
reparametrization invariance Ward
identities \CTrivi, and similarly for an arbitrary tachyon potential
\osdqm.  To lowest order in the potential strength, the critical
tachyon potential is a cosine with period $p_c$, and the theory with
this cosine
potential looks very much like the Sine-Gordon model restricted to a line.
In the Sine-Gordon model, the cosine potential that is at a zero of the
$\beta$-function to lowest order in $V_0$ no longer remains so at
higher orders in $V_0$.  However, in this paper we will prove that
in the 1-D theory the cosine potential that satisfies the $\beta$-function
to lowest order is at a renormalization group fixed point and has a
vanishing reparametrization invariance anomaly to all orders in the
potential.

The paper is organized as follows.  In Section~2 we give a brief review of
the open string boundary state and dissipative quantum mechanics.  In
Section~3 we present the reparametrization invariance Ward identities and
our regulation scheme.  In the following two sections,
we will evaluate the reparametrization Ward
identities for an arbitrary tachyon field and a constant,
uniform gauge field, diagrammatically.
The first part of Section~4 follows closely
the appendix of \osdqm.  The goal of these two
sections is to reduce the formal expression for the Ward identity to an
identity for the graphs in the perturbative expansion of the 1PI
function, summed to all orders in the string tension, $\alpha'$.
This result enables us to show that if the tachyon is periodic with
period $p_c$ (where $p_c$ depends on the strength of the gauge field),
and if we could ignore the regulator, then the
Ward identity would be satisfied to {\it all} orders in $\alpha'$.
Additionally, these calculations show that even when the regulator is taken
into account, the special choice of tachyon
potential does satisfy the Ward identity at every {\it finite} order in the
tachyon strength, $V_0$, and $\alpha'$.
However, others have shown \das\ that for closed string tachyons in 2D
sigma models, even if the $\beta$-function is satisfied to all finite orders
in $\alpha'$, new divergences arise once it is summed to all orders.  This
can give anomalous contributions to the $\beta$-function.  Therefore, in the
remainder of the paper we evaluate the graphs, summed over all loops, more
carefully.

In Section~6 we define what we mean by the flows generated by the Ward
identity, and in Section~7 we calculate the flow for the coupling constant
of $e^{i\vec q\cdot\vp(t)}$.  Readers who are only interested in the main
results may wish to skip sections 4, 5 (except for 5.5) and 7.
In Section~8 we find that, to
leading order in the cutoff, the Ward identity is equivalent to the
$\beta$-function.  This means that
the condition for the Ward identity to be satisfied is the same as
for the system to lie at a zero of the $\beta$-function.  Thus,
once the system is scale invariant, it does have the infinite set of
``hidden'' symmetries corresponding to the reparametrization
invariance of the underlying string theory.  Next, in Section~9, we discuss
the validity of our approximations, and in Section~10, we show that the flow
for the coupling constant of $d \phi(t)/dt$ is always zero.  In Section~11,
we show that, for special values of the gauge field, the cosine potential with
period $p_c$ does satisfy the Ward identity to all orders in $V_0$ and
$\alpha'$.  Consequently, the naive first order result is true to all
orders, and we have found exact fixed points of
dissipative quantum mechanics that correspond to solutions of string
theory.  In Section~12 we conclude the paper with a few remarks about the
consequences of our results.
\newsec{Background: Open String Theory and the Dissipative Hofstadter Model}

In this section, we briefly review the open string boundary state and its
connection to dissipative quantum mechanics.  For more details, the reader
is referred to references \dissqm, \osdqm, \cgcdef. (The first part of this
section closely follows the review in reference \cff).
In the presence of open string background fields,
interactions between a string and the background take place at the
boundary of the string. Their effects can be represented by a
boundary state $|B\rangle$, which lies in the closed string Hilbert space.

In \clny\ it is shown that this boundary state is given by
\eqn\bs{|B\rangle=\exp\left\{\sum_{m=1}^\infty{1\over m}\alpha_{-m}\cdot
          \tilde\alpha_{-m}\right\}
       \int\left[D\vX(s)\right]'
		\exp(-S_{KE} -S_{A}-S_{V} -S_{LS})|0\rangle,}
where
\eqn\SKEdef{S_{KE}={1\over8\pi^2\apm}\int_0^T ds\int_{-\infty}^\infty ds'
      {\bigl(\vX(s)-\vX(s')\bigr)^2\over(s-s')^2}~;}
\eqn\SAdef{S_A=i\int_0^T ds\,A_\mu(\vX)\dot X^\mu;}
\eqn\STdef{S_V= \int_0^T ds\,  V(\vX)~;}
and
\eqn\SLSdef{S_{LS}=\sqrt{2\over\alpha'}\int ds~\alpha(s)\cdot\vX(s)
      \qquad {\rm with}\quad
      \alpha^\mu(s)=\sum_{m=1}^\infty i(\tilde\alpha_{-m}^\mu e^{-ims}+
           \alpha_{-m}^\mu e^{ims})~.}
In these expressions, $T$ is the parameter length of the boundary,
$\apm$ is the string constant, and $A_\mu(\vX)$ and
$V(\vX)$ are the gauge fields and tachyon fields, respectively.
The creation operators of the left- and right-moving modes of the closed
string, $\tilde\alpha_{-m}$ and $\alpha_{-m}$, act on the closed string
vacuum $|0\rangle$ to create some state in the closed string Hilbert
space.  The notation $\bigl[D\vX(s)\bigr]'$
means that the zero-mode, $\vX_0$, is not integrated out. The commuting
objects $\tilde\alpha_{-m}$, $\alpha_{-m}$ and $\vX_0$ together make up a
set of coordinates which specify where the boundary lies in the target
space, and the boundary state is just a functional of these coordinates.
As an example of the usefulness of this construct, we note that the projection
of $|B\rangle$ onto the graviton state is essentially the energy-momentum
tensor of the open string object under study.  This gives us a
string-theoretic way to define such important notions as gravitational
and inertial mass.

This path integral is the generating functional for a renormalizable
``one-dimensional'' field theory described by the underlying action
$S_{KE}+S_A+S_V$. $S_{LS}$ is the linear source term in the generating
function.  This theory is divergent, so it requires a regulator.  One
possible choice is to add a dimension-two operator to the action of the
form
\eqn\SRdef{S_{R}=\int_0^T ds\,\half M\dot{\vX}^2(s)~.}
To obtain the boundary state, in the end we must take the cut-off, $M$, to
zero.  In order for this limit to be meaningful, the field theory must lie
at a renormalization group fixed
point, which implies that the gauge and tachyon fields must satisfy some
``vanishing beta function'' equations of motion for open string background
fields.  This means that the associated 1-D field theory must lie at a phase
transition.

This 1-D field theory also describes a quantum mechanical particle subject
to a potential, a magnetic field, and a dissipative force.  According to
the Caldeira-Leggett model \dissqm\ for dissipation, the particle interacts
linearly with a bath of harmonic oscillators.  The interaction strengths and
the frequencies of the oscillators are chosen so that in the classical limit
the particle is subject to ohmic dissipation.  Because the oscillators
appear only linearly and quadratically in the action, they can be integrated
out of the path integral.  The resulting action for the particle is
given by
\eqn\Sdqm{S_{dqm}= S_R+S_{KE}+S_A+S_V,}
where $S_R$, $S_{KE}$, $S_A$ and $S_V$ are defind in equations \SRdef,
\SKEdef, \SAdef, and \STdef.  However, now $M$ is the mass of the particle,
$A_\mu(\vX)$ is the vector potential and $V(\vX)$ is the scalar
potential.  The classical coefficient of friction, $\eta$, appears only
in $S_{KE}$ and is related to the string tension by $\eta = 1/(2\pi\apm)$.
In order to correspond to a solution of string theory, this system must lie
at a phase transition.

One dissipative quantum system which exhibits phase transitions is the
dissipative Wannier-Azbel-Hofstadter model (which is also known as the
dissipative Hofstadter model).  It is obtained by restricting
the particle to lie in two dimensions and taking
\eqn\DHMAdef{A_\mu(\vX)={eB\over 2c}\e_{\mu\nu}X^\nu}
and
\eqn\DHMVdef{V(\vX)= -V_0\cos({2\pi X(t)\over a})
                          -V_0\cos({2\pi Y(t)\over a}).}
This model depends on the dimensionless parameters defined by
\eqn\defalphat{2\pi\tilde \alpha = {\eta a^2 \over \hbar},\qquad
		2\pi\tilde \beta = {eB\over \hbar c} a^2~.}
In ref.\cgcdef\ we have shown that the phase diagram of this model should have
an infinite number of critical circles in the $\alpha-\beta$ plane, which
suggests that we have found many solutions to open string theory.

Motivated by this connection, in this paper we will only consider
the 1-D field theories where $\vX$ is restricted to lie in two dimensions
and the gauge field is taken to be a constant of the form given by
eqn. \DHMAdef. (Strictly speaking, to calculate the boundary state,
one must still include
the remaining 24 dimensions of $\vX$, which, for convenience, will be
assumed to be free fields.  Although there is some relation,
it is not clear that there is a direct connection between this model
with $\vX$ taken to be two-dimensional and open string
theory in 1+1 dimesions as considered, for example,
in reference \bkopen.)
Also in analogy with the dissipative quantum mechanics case, we would like
to define the dimensionless parameters as in equation \defalphat.  However,
for now we do not want to assume the tachyon field has a particular period,
so instead we will define the dimensionful parameters
\eqn\defalpha{\alpha={1\over \apm}, \qquad
              \beta = {2\pi eB\over c}.}
These differ from the usual choice of $\beta$ and $\alpha$ for the
dissipative Wannier-Azbel-Hofstadter model, given in eqn. \defalphat,
only in that we have left out the $\hbar$ and the factor of
$a^2/(2\pi)^2$.

\newsec{The Reparametrization Invariance Ward Identities}
 From the preceeding discussion, it might appear that we
have found many solutions to open string theory,
since the dissipative Wannier-Azbel-Hofstadter model has
many critical theories.  However, the space-time
physics of the string theory is not only invariant under changes in
scale, but it is also invariant under reparametrizations of the world
sheet.  This means that the boundary state must be invariant under
reparametrizations of the boundary of the world sheet.  The condition
for the reparametrization invariance of the boundary state is given by
\eqn\reparcond{\left(L_n-\tilde L_{-n}\right)|B\rangle=0
           \qquad{\rm for} \quad -\infty\le n\le\infty.}
Here, $|B\rangle$ is the boundary state defined in equation \bs,
and $L_n$ and $\tilde L_{-n}$ are the closed-string Virasoro
generators, given by
\eqn\Lndef{L_n={1\over2}\sum_{m=-\infty}^\infty\alpha_{n-m}\cdot\alpha_m,}
and similarly for $\tilde L_{-n}$.  For $-m<0$, the $\alpha_{-m}$ and
$\tilde\alpha_{-m}$
are the closed string creation operators; and for $m\ge0$, the $\alpha_m$
can be represented in terms of derivatives with respect to
the creation operators, $\alpha_{-m}$.  When we apply the operator
$L_n-\tilde L_{-n}$ to the expression for the boundary state in equation
\bs, then, as shown in \CTrivi, the condition for
reparametrization invariance is reduced to a set of Ward
identities. For $n\ge 0$, they are given by
\eqn\WIdef{\eqalign{0=\delta_n\Gamma
  &-\sum_{m=1}^{n-1}{1\over \alpha'}m(n-m)\phi_m\cdot\phi_{n-m}\cr
  &-\sum_{m=1}^{n-1}{1\over 2}m(n-m)
           {\partial^2W\over\partial\alpha_{-m}\cdot\partial\alpha_{m-n}}
   + in\sqrt{\alpha'\over2}{\partial^2W\over\partial X_0 \cdot
       \partial\alpha_{-n}} .}}
For $n<0$, we just replace $\alpha_{-k}$ with $\tilde\alpha_{-k}$ and
$\phi_k$ with $\phi_{-k}$.
In equation \WIdef, $W(\alpha, \tilde\alpha, \vec X_0)$ is the connected
generating functional given by
\eqn\Wdef{W(\alpha, \tilde\alpha, \vec X_0)= \int\left[D\vX(s)\right]'
		\exp(-S_{R}-S_{KE} -S_{A}-S_{V} -S_{LS})|0\rangle,}
and $\Gamma(\phi)$ is the 1PI generating functional.  It is obtained by
Legendre transforming W, as follows:
\eqn\gammadef{\Gamma(\phi^{cl}) = W(\alpha,\tilde \alpha,\vec X_0)
  -i\sqrt{2\over \alpha'} \sum_{m=1}^\infty(\alpha_{-m}\cdot\phi_m^{cl}
                   +\tilde\alpha_{-m}\cdot\phi_{-m}^{cl}),}
where $\phi^{cl}$ and $\alpha$ are related by
\eqn\phicldef{\eqalign
{\phi_m^{cl}&= -i\sqrt{\alpha'\over2}{\partial W \over \partial\alpha_{-m}}
\qquad
\phi_{-m}^{cl}= -i\sqrt{\alpha'\over2}
               {\partial W \over \partial\tilde\alpha_{-m}} \cr
\alpha_{-m}&= i\sqrt{\alpha'\over2}
             {\partial\Gamma \over \partial \phi_m^{cl}} \qquad
\tilde\alpha_{-m} = i\sqrt{\alpha'\over2}
        {\partial\Gamma \over \partial \phi_{-m}^{cl}} \cr
&\qquad \qquad \phi_0^{cl} = X_0. }}
The first term in equation \WIdef, $\partial_n\Gamma$,
is the variation of $\Gamma$ under an infinitesimal reparametrization
$s \rightarrow s +  f_n$, with $f_n(s) = i \epsilon e^{ins}$.  The
remaining terms reflect the fact that the effective action is not
reparametrization invariant.  In fact, the non-local kinetic term, $S_{KE}$,
in the classical action is
not reparametrization invariant, so the second term in the Ward identity
subtracts off its variation, and the third term corrects for the effect that
this non-invariance has on higher order diagrams.  The last term
makes up for the fact that we have selected out a particular
parametrization in order to define the zero mode of $\vec X$ so that we could
treat it separately.

To evaluate the Ward identity, we use a perturbative background-field
expansion.  To find $\Gamma(\phi^{cl}(t))$, we begin by expanding $V(\vX)$
around $\vX(t) = \vec\phi^{cl}(t)$, where $\vec\phi^{cl}_m$ is defined in
equation \phicldef.  We will assume that $\vec\phi^{cl}(t)$ is a smooth
function of $t$, and we will use the notation $\vec \phi^{cl}(t)$ and
$\vec \phi(t)$ interchangeably.
If we set $\vec X=\vec \phi+\vec x$, then $S_V$ becomes
\eqn\SVtay{S_V = \int {ds\over 2\pi}V(\vec X)
               = \int{ds\over 2\pi} V(\vec \phi) +
              \int{ds\over 2\pi} \nabla\mu V(\vec \phi)x^\mu +
{1\over 2}\int{ds\over 2\pi} \nabla\mu \nabla\nu V(\vec \phi)x^\mu x^\nu
+ \dots}
Similarly, to find $W(\alpha, \tilde\alpha, \vec X_0)$, we expand
$V(\vec X)$ around the
zero mode, $\vec X_0$.  Next, we treat $S_V$ as a perturbation and Taylor
expand $\exp\left(-\int S_V\right)$.  The propagator is then determined by
the remaining quadratic terms in the action, $S_{KE}+S_A$ and $S_R$.
However, as explained above, $S_R$ acts only as an ultra-violet regulator,
so we will set $M$ to zero and choose a more convenient cutoff.  The
quadratic part of the action can then be written in momentum space as
\eqn\fsact{S_{KE}+S_A= {1\over2}\int{d\omega \over 2\pi}
                        \tilde X_\mu^\dagger(\omega) \tilde X_\nu(\omega)
                        \tilde S^{\mu\nu}(\omega),}
where the inverse propagator, $\tilde S(\omega)$, is given by
\eqn\invprop{\tilde S^{\mu\nu}(\omega)
      ={\alpha\over 2\pi}|\omega|\delta^{\mu\nu}
       + {\beta\over 2\pi}\omega\epsilon^{\mu\nu}.}
For the ultra-violet regulator, we will multiply the propogator by an
$\exp\left(-\epsilon |\omega|\right)$ cutoff.  In addition, we will need an
infra-red cutoff, which we will define by putting the time, $t$, on a circle
of circumference $T$.  (For simplicity, we will take $T=2\pi$, but this
should not significantly affect our results.)
The diagonal part of the propagator is then given by
\eqn\Gmmdef{\eqalign{G^{\mu\nu}(t)
          =&\aab \sum_{m\neq0}{1\over |m|}e^{imt}
          e^{-\epsilon|m|} \delta^{\mu\nu}\cr
        =&-\aab \ln\left(1 + e^{-2\epsilon}-2e^{-\epsilon}
            \cos t\right)\delta^{\mu\nu}. }}
The off-diagonal part is
\eqn\Gmndef{\eqalign{G^{\mu\nu}(t) =&-\epsilon^{\mu\nu}\bab
           \sum_{m\neq0}{1\over m}e^{imt} e^{-\epsilon|m|}\cr
          =&\epsilon^{\mu\nu} \bab
    \ln \left[{\sin\left\{(t+i\epsilon)/2\right\} \over
           \sin\left\{(-t+i\epsilon)/2\right\}} e^{it} \right]. }}
Note that when $\beta=0$, we have $\aab = \alpha'$.

We will find that
all diagrams of interest can be written in terms of the following
sum over $G^{\mu\nu}(t)$:
\eqn\Emndef{E(t;\mu,\nu) = \sum{1\over n!} (G(t)^{\mu\nu})^n
                         = e^{G^{\mu\nu}}.}
The propagator $G^{\mu\nu}(t)$, and not the propagator $E(t;\mu,\nu)$,
determines the connectedness and irreducibility of the diagrams.
That means that if we express the graphs in terms of $E(t;\mu,\nu)$,
it is more difficult to keep track of whether they are connected or 1PI.
However, by using the propagator $E(t; \mu,\nu)$, we will have automatically
summed over all orders in $\alpha'$ or $1/\alpha$.

\newsec{Ward Identity with $\beta = 0$}
We will begin by considering the case when $\beta = 0$ and restrict
$\vec X=x$ to lie in one dimension.  In that case, according to standard
perturbation theory for path integrals, the general $O(V_0^N)$ graph
will have $N$ vertices with a factor of $V(x(t_i))$ associated to
each vertex.  The $i^{\rm th}$ vertex has $M_i$ legs coming out of it,
and for each leg there is a derivative ${\partial \over \partial x(t_i)}$
acting on $V(x(t_i))$.  Each leg is joined to one other leg by the
propagator $G(t)$, which is obtained from $G^{xx}(t)$ by setting $\beta = 0$.
We will define a petal to be a propagator that joins two legs from the same
vertex, and we will let $P_i$ equal the number of petals coming out of the
$i^{\rm th}$ vertex.  Also, we will define a link to be a propagator that
joins legs coming from two different vertices, and we will let $l_{ij}$
be the number of links between two different vertices $i$ and $j$.
The total number of legs coming from a vertex is related to the number
of petals and links coming from the vertex by the equation
\eqn\MPlrel{M_i = 2P_i + \sum_j l_{ij}.}
For example, the following graph with three vertices has $P_1=0$,
$P_2=3$, $P_3=2$, $l_{12}=1$, $l_{23}=1$ and $l_{31}=2$.
\eqn\gengraphdiag{\left[{\partial\over\partial x(s_1)}\right]^3
                  \left[{\partial\over\partial x(s_2)}\right]^8
                  \left[{\partial\over\partial x(s_3)}\right]^7\,\,
                  \vcenter{\epsffile{Wione.eps}}.}
The value of a general graph is given by
\eqn\gengraph{\eqalign{\Gamma(l_{ij},P_i) =-&\nf \symf\cr
                &\quad\times\siint\gpi\glij
                 \prod_i\left[\part{x(s_i)}\right]^{M_i}V(x(s_i)).}}
The symmetry factor is exactly what one would expect.  $1/\lij!$ takes into
account that there are $\lij$ identical links joining the $i^{\rm th}$ and
$j^{\rm th}$ vertex, and $1/(P_i!2^{P_i})$
takes into account that there are $P_i$
identical petals coming from each vertex, each with two identical end points.
To make a graph for the 1PI vacuum function, we just replace $x(s_i)$ with
$\phi^{cl}(s_i)$ (we will write $\phi(s_i)$ for short), and
keep only those graphs that remain connected when any one link is cut.
For the connected generating-functional, we must remember that the action
includes the linear source term
$i\sqrt{2/\alpha'} \sum_{m\ne 0} \alpha_m x_m$,
which generates extra vertices weighted by $\sqrt{2/\alpha'} \alpha(t)$ and
having only one leg.  Then the $O(V_0^N)$ graph for $W$ is given by replacing
$x(s_i)$ by $X_0$, the zero mode, and keeping
only the connected graphs of the form
\eqn\Wdef{\eqalign{W_{l_{ij},P_i} (\alpha) =
     -&\nf\symf\prod_{i\leq N}\left[\part{X_0}\right]^{M_i}V(X_0)\cr
     &\quad\times\siint \left[G(0)\right]^{\sum_{i=1}^N P_i} \glij
     \prod_{i>N}\sqrt{2\over \alpha'}\alpha(s_i).}}
An example of such a graph is
\eqn\Wdiag{\left(\partial\over\partial X_0\right)^3V(X_0)
           \left(\partial\over\partial X_0\right)^5V(X_0)
           \left(\partial\over\partial X_0\right)^8V(X_0)\,\,
           \vcenter{\epsffile{Witwo.eps}}.}

\subsec{Variation of the 1PI Vacuum Function}
First, we will calculate $\delta_n \Gamma(\phi(s))$. Strictly speaking, this is
the change in $\Gamma(\phi(s))$ when $s \rightarrow s + f_n$,
where $f_n$ is given by $f_n(s) = i\epsilon e^{ins}$.  Under this
reparametrization, the change in $\phi$ is
$\delta\phi(s) = \left[{\partial\phi \over \partial s}\right] f_n(s)$,
which implies that the change in $V(\phi)$ is given by
\eqn\dV{\delta V(\phi(s_i)) =
       \left[{d\over d\phi(s_i)} V(\phi(s_i))\right]\delta\phi
       = \left[{d\over ds_i} V(\phi(s_i))\right] f_n(s_i).}
Then $\delta_n \Gamma(\phi(s))$ is just the sum over graphs for $\Gamma$
with one of the $V(\phi(s_i))$'s replaced by
$f_n(s_i){d\over ds_i} V(\phi(s_i))$.  Now we can integrate by parts.  For each
vertex, we obtain two types of terms, one with the derivative acting on
$f_n(s_i)$, and the other with the derivative acting on $G(s_i-s_j)$. The
latter is just the change in $G$ under a reparametrization of $s_i$.  The
resulting graphs are minus the change of $\Gamma$ under
reparametrizations where $V(\phi(s_i))$ behaves like a scalar.  The first
term, when $f_n(s_k)$ is varied, is given by
\eqn\defhs{\eqalign{\delta_n\Gamma_P(\lij,P_i)
            = &\nf\symf\gpi\cr
            &\times\siint
            f_n'(s_k) \glij\prod_i\left[\part{\phi(s_i)}\right]^{M_i}
            V(\phi(s_i)) \cr
            = &\int {ds_k\over 2\pi} f_n'(s_k) h_{\lij,P_i}(s_k).}}
In this equation, $h_{\lij,P_i}(s_k)$ is defined to be a graph that is
integrated over all vertices except for $s_k$, with $P_i$ petals at the
$i$th vertex and $l_{ij}$ links joining the $i$th and $j$th vertex.
The second term, when $G(s_k-s_m)$ is varied, is given by
\eqn\defhss{\eqalign{\delta_n\Gamma_L(\lij,P_i) = &\nf\symf\siint
            f_n(s_k) \gpi\cr
            &\qquad\times\glij l_{km} {{d\over ds_k}G(s_k-s_m)\over G(s_k-s_m)}
            \prod_i\left[\part{\phi(s_i)}\right]^{M_i}V(\phi(s_i))\cr
            = &\int{ds_k\over 2\pi}{ds_m\over 2\pi}f_n(s_k)
              \left[{d\over{ds_k}}G(s_k-s_m)\right]\part{\phi(s_k)}
              \part{\phi(s_m)} h_{\lij,Pi}(s_k, s_m).}}
Here, we have defined $h_{\lij,Pi}(s_k, s_m)$ to be a graph that is
integrated over all vertices except for $s_k$ and $s_m$.  It has $P_i$
petals at the $i$th vertex; $l_{ij}$ links joining the $i$th and $j$th
vertex for $\{i,j\} \ne \{k,m\}$; and $l_{km}-1$ links joining the $k$th and
$m$th vertex.
The two types of graphs we obtain from varying
$\Gamma(l_{ij}, P_i)$ can be summarized by
\eqn\onepidiag{\delta_n\Gamma(\lij,P_i)
        = \quad\hbox{\lower.33in\hbox{\epsffile{Withree.eps}}}
      \quad + {\partial\over\partial\phi(s_k)}{\partial\over\partial\phi(s_m)}
      \,\,\hbox{\lower.33in\hbox{\epsffile{Wifour.eps}}},}
where
\eqn\hsdiag{\hbox{\lower.33in\hbox{\epsffile{Wifoura.eps}}}\,\,
                =h_{l_{ij},P_i}(s_k),}
\eqn\hssdiag{\hbox{\lower.33in\hbox{\epsffile{Wifourb.eps}}}\,\,
                =h_{l_{ij},P_i}(s_k, s_m),}
and
\eqn\dGssdiag{\vcenter{\epsffile{Wifourc.eps}}\,\,
              = {d\over d s_k}G(s_k,s_m).}

The second term in the Ward identity explicitly subtracts off the variation of
the part of $\Gamma$ due to the classical non-local kinetic term.

\subsec{Evaluation of the ``Kinetic Term'' in the Ward Identity}
Now we turn our attention to the third term in the Ward identity.  We
will need the identity (used in \osdqm)
\eqn\hHid{{1\over \alpha'}\sum_{m=1}^{n-1}m(n-m) \tilde F_{m,n-m} =
           {1\over 2} \int{ds\over 2\pi}{ds'\over 2\pi}\delta_f
            H_0(s-s')F(s,s'),}
where $F$ is an arbitrary function of $s$ and $s'$, and $\tilde F$ is its
Fourier transform.  $H_0$ is the inverse propagator,
\eqn\Hzdef{H_0(s-s') = \sum_{-\infty}^\infty {|m| \over\alpha'} e^{im(s-s')},}
and $\delta_fH_0$ is given by
\eqn\dfHzdef{\delta_fH_0 = {d\over ds}\left[f_n(s)H_0(s-s')\right]
                                    -H_0(s-s') f_n(s'){d\over ds'}.}
Then, the third term in the Ward identity can be written as
\eqn\Ddef{D = \int{ds\over 2\pi}{ds'\over 2\pi} \delta_fH_0(s-s')
                  {\partial^2W\over\partial\alpha(s)\partial\alpha(s')}.}
%
This expression can be viewed as coming from the connected diagrams
where $\alpha(s)$ and $\alpha(s')$ have been amputated and then the two
legs are joined by $\delta_fH_0(s-s')$ instead.  For example, if we
perform this operation on the graph in \Wdiag, we obtain
\eqn\Dconndiag{\left(\partial\over\partial X_0\right)^3V(X_0)
           \left(\partial\over\partial X_0\right)^5V(X_0)
           \left(\partial\over\partial X_0\right)^8V(X_0)\,\,
           \vcenter{\epsffile{Wifive.eps}}.}
Alternatively, when
the expression is Legendre transformed, $D$ comes from all 1PI diagrams
with $\delta_fH_0$ inserted on a propagator.  (To perform the Legendre
transformation, one can start with any such
``decorated'' 1PI diagram and obtain the corresponding connected
diagrams by first expanding each vertex factor, $V(\phi^{cl}(s))$, around
$\phi^{cl}(s) = x_0$ and then replacing $\phi^{cl}(s)$ with the diagrams for
${\partial W\over\partial\alpha}$.)  The contribution to $D$ when
$\delta_f H_0$ is inserted between $s_1$ and $s_2$ in the 1PI diagram
in \gengraphdiag\ is
\eqn\Donepidiag{\eqalign{\left[{\partial\over\partial \phi(s_1)}\right]^3
                  &V\left(\phi(s_1)\right)
                  \left[{\partial\over\partial \phi(s_2)}\right]^8
                   V\left(\phi(s_2)\right)
                  \left[{\partial\over\partial \phi(s_3)}\right]^7
                   V\left(\phi(s_3)\right)\cr
                 \noalign{\medskip}
                 &\qquad\times\vcenter{\epsffile{Wisix.eps}}.}}

When $\delta_f H_0(s-s')$ is inserted into a propagator in a 1PI graph, there
are two different cases, represented by
\eqn\DdHdiag{D=-{1\over2}{\partial^2\over\partial\phi^2(s)}\,\,
             \hbox{\lower.33in\hbox{\epsffile{Wiseven.eps}}}\,\,
        -{\partial\over\partial\phi(s)} {\partial\over\partial\phi(t)}
         \,\,\hbox{\lower.33in\hbox{\epsffile{Wieight.eps}}},}
where
\eqn\GHdefdiag{\vcenter{\epsffile{Wieighta.eps}}=G(s-t)\qquad {\rm and}
          \quad\vcenter{\epsffile{Wieightb.eps}}=\delta_fH_0(s'-s").}
The first graph is due to inserting $\delta_fH_0$ on a petal (which
is calculated in reference \osdqm) and the second, to inserting it on a link
(which is discussed in reference \larsthesis).  In both cases we will need the
convolution of $H_0$ with $G$, which ought to be a delta function.
However, because we are regulating $G$ and excluding the sum over the
zero mode in $G$, the convolution is given by
\eqn\HGint{\int{ds''\over2\pi} H_0(s'-s'')G(s''-s) = \de(s'-s)-1,}
where
\eqn\defdelt{\Delta_\epsilon(s-s') = \sum_{m=-\infty}^\infty e^{im(s'-s)}
                                      e^{-\epsilon|m|}
                             = {\sinh\epsilon\over \cos\epsilon-\cos(s-s')},}
and $\Delta_\epsilon(s'-s) \rightarrow \delta(s-s')$ as
$\epsilon\rightarrow 0$.

When $\delta_f H_0(s-s')$ is inserted in a petal, we must evaluate the
following integral:
\eqn\Dpdef{D_P(\lij,P_k)=-{1\over2}\int{ds\over 2\pi}{ds'\over 2\pi}
        {ds''\over 2\pi}{\partial^2\over\partial\phi^2(s)} h_{\lij,P_k}(s)
        G(s-s')\delta_fH_0(s'-s'')G(s''-s),}
where $h_{\lij,P_k}(s)$, as defined by eqn. \defhs, is a 1PI graph
integrated over all vertices except $s$.  Similarly, when $\delta_fH_0(s-s')$
is inserted into a link, the graphs we obtain have the following value:
\eqn\Dldef{\eqalign{ D_L(\lij,P_k) = -\int \dtp s&\dtp t\dtp{s'} \dtp{s''}
           \part{\phi(s)}\part{\phi(t)} h_{\lij,P_k}(s,t)\cr
           &\times G(s-s')\delta_fH_0(s'-s'')G(s''-t),}}
where $h_{\lij,P_k}(s,t)$ is a 1PI graph integrated over all vertices
except $s$ and $t$, as defined in eqn. \defhss.  For both of these integrals,
we need to evaluate
\eqn\Iintdef{I=\int\dtp{s'}\dtp{t'} G(s-s')\delta_fH_0(s'-t')G(t'-t),}
where $s$ and $t$ can be equal.  Using the definition of $\delta_fH_0$ and
integrating by parts, we have
\eqn\fullI{\eqalign{I=&-\int\dtp{s'}\dtp{t'}\left({d\over ds'}G(s-s')\right)
              f_n(s')\left[H_0(s'-t')G(t'-t)\right]\cr
                    &-\int\dtp{s'}\dtp{t'}\left[G(s-s')H_0(s'-t')\right]
              f_n(t'){d\over dt'}G(t'-t).}}
When we perform the integral over $H_0G$, we obtain the function $\Delta-1$,
an approximate delta-function minus a
constant term due to the zero mode.  The part of
$I$ resulting from the constant term is
\eqn\Icdef{I_c = \int\dtp{s'} f_n(s') \left[{d\over ds'}G(s-s')+
                 {d\over ds'}G(t-s')\right].}
Upon integrating by parts, it becomes
\eqn\finalIc{I_c = -\int\dtp{s'} f_n'(s')\left[G(s-s')+G(t-s')\right].}
The terms in $I$ coming from the approximate delta-function have the form
\eqn\Ideltdef{I_\Delta=-\int\dtp{s'}f_n(s')\left[\de(t-s')
                  {d\over ds'}G(s-s')+\de(s-s'){d\over ds'}G(t-s')\right].}

Returning to the integrals for the full diagrams, when $\delta H$ is inserted
into a petal, we have two integrals.  The first is due to ignoring the
zero mode in $G$, and it is obtained from $I_c$ when $s=t$.  It is given by
\eqn\DPzdef{D_{P,0}(\lij,P_k)= \int\dtp s\dtp{s'}
           {\partial^2\over\partial \phi^2(s)}\hs s f_n'(s')G(s-s').}
The second term comes from the approximate delta function and is obtained from
$I_\Delta$ with $s=t$.  It has the form
\eqn\DPdeldef{\eqalign{D_{P,\Delta}(\lij,P_k)=&\int\dtp s\dtp{s'}
           {\partial^2\over\partial \phi^2(s)}\hs s f_n(s')
           \Delta_\epsilon(s-s'){d\over ds'}G(s-s')\cr
    =&{\partial^2\over\partial\phi(s)}\,\,
       \hbox{\lower.33in\hbox{\epsffile{Witen.eps}}}.}}
We are interested in the limit as $\epsilon \rightarrow 0$.  In that limit,
$\Delta_\epsilon$ becomes a delta function, so, for small $\epsilon$,
we should be able to evaluate the integral for small values of the argument
of $\Delta_\epsilon$.  For small $\epsilon$ and $t$, $\de$ and $G$ become
\eqn\smalldelta{\de(t) \approx {2\epsilon\over \epsilon^2+t^2}
                        (1 + O(\epsilon^2)+O(t^2))}
and
\eqn\smallG{{d\over dt}G(t) \approx -{2\alpha' t\over t^2 + \epsilon^2}
                        ((1 + O(\epsilon^2)+O(t^2))}
To perform the integral, we will also make the change of variables
\eqn\chvar{s_+ = s \qquad s_- = s-s';}
and we Taylor expand $f$ around $s_+$, keeping the lowest order
term that gives a non-vanishing contribution to $D_{P,\Delta}$.
Then we obtain the following expression for
$D_{P,\Delta}$:
\eqn\Dpdelint{D_{P,\Delta}(\lij,P_k)= -\int\dtp {s_+}\dtp{s_-}
           {\partial^2\over\partial \phi^2(s_+)}\hs{s_+} f_n'(s_+)
           {4\alpha'\epsilon s_-^2\over(\epsilon^2 + s_-^2)^2}.}
After integrating over $s_-$, we get the final form for $D_{P,\Delta}$,
\eqn\finalDpdelta{D_{P,\Delta}(\lij,P_k)= -\alpha'\int\dtp {s_+}
           {\partial^2\over\partial \phi^2(s_+)}\hs{s_+} f_n'(s_+).}
As a result, the two types of graphs involved in the calculation of
$D_P$ are
\eqn\DPdiag{D_P(l_{ij},P_k)=-{\partial^2\over\partial^2\phi(s)}\,\,
          \hbox{\lower.33in\hbox{\epsffile{Winine.eps}}}\,\,
         -\alpha'{\partial^2\over\partial^2\phi(s_+)}\,\,
          \hbox{\lower.33in\hbox{\epsffile{Wieleven.eps}}}\,\,.}

Now we must calculate the effect of inserting $\delta_fH_0$ into a link.
Again we have two integrals.  The first is
\eqn\DLzdef{D_{L,0}(\lij,P_k) = {1\over2}\int\dtp s\dtp t\dtp s'
                       \part{\phi(s)}\part{\phi(t)}\hst st
                       f_n'(s')\left[G(s-s')+G(t-s')\right];}
and the second is
\eqn\DLdeltdef{\eqalign{D_{L,0}(\lij,P_k) = {1\over2}&\int\dtp s\dtp t\dtp s'
            \part{\phi(s)}\part{\phi(t)}\hst st f_n(s') \cr
            &\times\left[\de(t-s'){d\over ds'}G(s-s')+
            \de(s-s'){d\over ds'}G(t-s')\right].}}
Thus, the graphs for $D_L$ are represented by
\eqn\DLdiag{\eqalign{{1\over2}{\partial\over\partial\phi(s)}
                     {\partial\over\partial\phi(t)}&
 \left[\,\,\hbox{\lower.33in\hbox{\epsffile{Witwelve.eps}}}\,\,\right]\cr
           + {1\over2}{\partial\over\partial\phi(s)}
                     {\partial\over\partial\phi(t)}&
 \left[\,\,\hbox{\lower.33in\hbox{\epsffile{Withirteen.eps}}}\,\,\right]
  .}}

\subsec{Evaluation of the Zero-Mode Term in the Ward Identity}
Finally, we must evaluate the last term in the Ward Identity.  It is
given by
\eqn\Cdef{C = in\sqrt{\alpha'\over2}{\partial^2W\over\partial X_0
                  \partial\alpha_{-n}}
            =-\int\dtp t f_n'(t)\part{X_0}{\partial W\over\partial\alpha(t)},}
where ${\partial W\over\partial\alpha(t)}$ is the sum of all the connected
graphs with one $\alpha$ truncated, leaving a free leg; and where
$\part{X_0}$ acts on the $V(X_0)$'s.  Again we have two different situations.
The first is when the $\part{X_0}$ acts on the same vertex that the free leg
is connected to, and the second is when the $\part{X_0}$ acts on a
different vertex.  If we calculate the contribution to $C$ from the
graph in \Wdiag, then in the first case we will get graphs like
\eqn\Csdiag{\vcenter{\epsffile{Wifourteen.eps}}\,\,,}
and in the second case, graphs like
\eqn\Codiag{\vcenter{\epsffile{Wififteen.eps}}\,\,.}
More generally, we can write
\eqn\cccid{\eqalign{C(\lij,P_k) =&C_1(\lij,P_k)  + C_2(\lij,P_k)\cr
    =&\,\,\hbox{\lower.33in\hbox{\epsffile{Wisixteen.eps}}}\,\,
    +\,\,\hbox{\lower.33in\hbox{\epsffile{Wiseventeen.eps}}}\,\,
    ,}}
where
\eqn\Conedef{C_1(\lij,P_k) = \int\dtp t\dtp s f_n'(t)G(t-s)\hsa s}
and
\eqn\Ctwodef{C_2(\lij,P_k) = \int\dtp t\dtp s\dtp{s'} f_n'(s')G(s'-s)\hsta st.}
$\hsa s$ is a connected graph integrated over all its vertices, except $s$,
and it has two extra derivatives, ${\partial^2 \over\partial X_0^2}$ acting
on the vertex at $s$.  For example, for the graph in \Csdiag, $\hsa{t_1}$
is given by
\eqn\hsadiag{\hsa{t_1} = \vcenter{\epsffile{Witwenty.eps}}.}
Similarly, $\hsta st$ is a connected graph integrated
over all vertices except $s$ and $t$, and it has two extra $\part{X_0}$'s, one
acting on the $V(X_0)$ at vertex $s$ and one on the $V(X_0)$ at vertex $t$.

In order to compare $C_1$ with $D_{P,0}$ and $C_2$ with $D_{L,0}$
we must perform the Legendre transformation on $D$ that is described after
equation \Dconndiag.  From this transformation, we obtain all
the connected graphs $\hsat s$ from
${\partial^2\over\partial\phi^2(s)} \hs s$ and all the connected
graphs $\hstat st$ from $\part{\phi(s)}\part{\phi(t)}\hst st$.
In particular, for a connected graph with petals labeled by $P_k$
and links labeled by $\lij$, we will obtain a contribution to
$D$ of the form
\eqn\finalDPz{D_{P,0}(\lij, P_k) = \int\dtp s\dtp t f_n'(t)G(s-t)\hsa s}
and
\eqn\finalDLz{D_{L,0}(\lij, P_k) = \int\dtp s\dtp t\dtp{s'}
                                    f_n'(s')G(s-s')\hsta st.}
As a result, when we subtact the third term, $D$, from the last term, $C$,
in the Ward identity, $D_{P,0} +D_{L,0}$ cancels with $C_1+C_2$, (after
appropriately symmetrizing in $s$ and $t$).  As we expected, the terms
in the Ward identity arising from the separate treatment of the zero
mode are accounted for by the last term in the Ward identity.

\subsec{Summing the Graphs}
The remaining terms in the Ward identity come from $\delta_n\Gamma$ and
$D_{P,\Delta}+D_{L,\Delta}$.  For a graph described by $\{\lij, P_k\}$, the
terms in the Ward identity due to varying a petal at the vertex labeled
by s are given by
\eqn\Alpsdef{\eqalign{A(\lij,P_k;s)
                 =& \delta_n\Gamma_P(\lij,P_k)-D_{P,\Delta}(\lij,P_k)\cr
                 =& \int\dtp s f_n'(s)(1+\alpha'\dal_s)\hs s\cr
                 =&(1+\alpha'\dal_s)\,\,
             \hbox{\lower.33in\hbox{\epsffile{Wieighteen.eps}}}
              \,\,,}}
where $\dal_s = {\partial^2\over\partial\phi^2(s)}$.
Similarly, the contribution to the Ward identity from the same graph when
a link between vertices labeled by s and t is varied is

\eqn\Alpstdef{\eqalign{A(\lij,P_k;s,t)
=&\delta_n\Gamma_L(\lij,P_k)-D_{L,\Delta}(\lij,P_k) \cr
                     =& \int\dtp s\dtp t \part{\phi(s)}\part{\phi(t)}\hst st\cr
  &\Biggl\{f_n(s){d\over ds}G(s-t)+f_n(t){d\over dt}G(s-t)\cr
  &-\int\dtp s' f_n(s')\left[\de(t-s'){d\over ds'}G(s-s')
                        +\de(s-s'){d\over ds'}G(t-s')\right]\Biggr\}.}}
(To obtain this, we have symmetrized in $s$ and $t$.)
The contribution due to varying a link can be summarized
diagrammatically as:
\eqn\Alpstdiag{\eqalign{\delta_n\Gamma_L&(\lij,P_k)-D_{L,\Delta}(\lij,P_k)\cr
    \noalign{\medskip}
    =&{\partial\over\partial\phi(s)} {\partial\over\partial\phi(t)}
    \,\,\hbox{\lower.42in\hbox{\epsffile{Winineteena.eps}}}\cr
    \noalign{\medskip}
    &\qquad\qquad\qquad\qquad
     \hbox{\lower.32in\hbox{\epsffile{Winineteenb.eps}}}
     \,\, .}}

These equations were only for the contribution of $\delta H$ inserted into
a particular petal or link of a particular graph.  We must sum over all
locations of the insertion of $\delta H$ and then over all 1PI graphs.
First we define
\eqn\WPdef{W_P(\lij,P_k) =\sum_s A(\lij,P_k; s),}
where $s$ runs over all the vertices in the graph, and
\eqn\WLdef{W_L(\lij,P_k) =\sum_{\{s,t\}}A(\lij,P_k; s,t),}
where the sum is over all pairs of vertices, $\{s,t\}$, in the graph.
Next, to perform the sum over all 1PI graphs with $N$ vertices, we find
it much more convenient to extend the sum to one over all graphs with $N$
vertices and then later subtract off the disconnected and one-particle
reducible graphs.  In that case, the $\lij$'s and $P_k$'s are allowed to
range from zero to infinity, and we are interested in the expressions
\eqn\hlijsum{\sum_{\{\lij\}=0}^\infty\sum_{\{P_k\}=0}^\infty \hs{s_m}
       \quad{\rm and}\quad\sum_{\{\lij\}=0\atop \lij\ne l_{mn}}^\infty
         \sum_{l_{mn}=1}^\infty\sum_{\{P_k\}=0}^\infty
          h_{\tilde\lij,P_k}(s_m,s_n).}
$\hs {s_m}$ and $h_{\tilde\lij,P_k}(s_m,s_n)$ are both integrals over
\eqn\hintegrand{h_{l_{ij},P_k}(\phi)
                =\nf\symf\gpi\glij\prod_i\left[\part{\phi(s_i)}\right]^{M_i}
                 V(\phi(s_i)),}
where $M_i = \sum_j(2P_j+\lij)$, and $\tilde\lij = \lij$ except for
$\tilde\l_{mn}$, which equals $l_{mn}-1.$
We can easily sum this expression over the $P_i$'s and $\lij$'s to obtain
\eqn\finalh{h(\phi)
               = \nf\prod_{i<j}e^{G(s_i-s_j)\part{\phi(s_i)}\part{\phi(s_j)}}
                \prod_i e^{{1\over2}G(0)\dal_i}(V(\phi(s_i))).}
Therefore, when we sum $W_P(\lij,P_k)$ and $W_L(\lij,P_k)$ over the $\lij$'s
and $P_k$'s, the Ward identity at order $V^N$ becomes
\eqn\WPWLZ{W_p + W_L=0,}
where
\eqn\WPdef{\eqalign{W_P=&\sum_{\{\lij\}=0}^\infty\sum_{\{P_k\}=0}^\infty
                      W_P(\lij,P_k)\cr
                     =&\nf\sum_k\siint f'(s_k)[1 +\alpha'\dal_k] h(\phi)}}
and
\eqn\WLdef{\eqalign{W_L=& \sum_{\{\lij\}=0\atop \lij\ne l_{mn}}^\infty
         \sum_{l_{nm}=1}^\infty\sum_{\{P_k\}=0}^\infty W_L(\lij,P_k)\cr
                     =&\nf\sum_{k\ne m}\siint \part{\phi(s_k)}
                        \part{\phi(s_m)}h(\phi)\cr
                     &\times\left[f(s_k){d\over ds_k}G(s_k-s_m)
          -\int dt\,f(t)\left[\de(t-s_k){d\over dt}G(s_m-t)\right]\right] .}}

\newsec{The Ward Identity with Uniform Magnetic Field}
When a non-zero magnetic field is applied, there are two changes
to our system.  The first is that $\vec \phi(s)$ and $\vec X_0$
now have two components.  We will call $\phi^x(s)=x(s)$ and
$\phi^y(s)=y(s)$.
The second change is that $G(s)$ is replaced by $\gmn(s)$.
The diagonal term in $\gmn$ is just $G(s)$ rescaled:
\eqn\gmmaabg{\gmm(s) = {\alpha^2\over\alpha^2+\beta^2}G(s).}
The off-diagonal term of $\gmn$ is given by equation \Gmndef.
A sample 1PI graph is
\eqn\samplediag{\vcenter{\epsffile{Magone.eps}},}
where
\eqn\GxxGxydiag{\vcenter{\epsffile{Magonea.eps}}=\gxx(s-t)
         \qquad {\rm and} \quad
                \vcenter{\epsffile{Magoneb.eps}}=\gxy(s-t).}
For simplicity, we will assume that at each vertex all the legs are
$x$'s or all are $y$'s.  This is the case for the cosine potential of
the dissipative Hofstdater model in equation \DHMVdef.  The results of this
section are not changed if we take a more general potential instead.

The calculation of the Ward identity proceeds in much the same way as
when $\beta=0$.  Apart from the rescaling of the diagonal propagator,
the only changes occur when an off-diagonal propagator is involved in some
way.  It turns out that the derivative of the off-diagonal propagator and
its convolution with $H_0$ are related to similar functions of the
original propagator, $G(s)$.  In fact, they satisfy the following
identities.
\eqn\ddsGxy{\eqalign{\der{s'}\gxy(s-s') =&i\bab\left[\de(s-s')-1\right]\cr
                      =&i\bab\int\dtp{s''}H_0(s'-s'')G(s''-s)}}
and
\eqn\HGxy{\int\dtp{s''}H_0(s'-s'')\gxy(s-s'') =
           -i{\alpha^2\beta\over\alpha^2+\beta^2}\der{s'}G(s-s').}

\subsec{Variation of the 1PI Vacuum Function}
We will now proceed to evaluate the diagrams, and again begin with
$\delta_n\Gamma$.  This time, when we reparametrize the 1PI diagrams,
we obtain a term due to the reparametrization of the $ds$,
\eqn\dgammaP{\delta_n\Gamma_P(\lij,P_k) = \int\dtp s f_n'(s)\hs s ;}
and a term due to the reparametrization of $\gmn(s-t)$,
\eqn\dgammaL{\delta_n\Gamma_L(\lij,P_k) = \int\dtp s \dtp t f_n(s)
            \left[\der{s}\gmn(s-t)\right]
            \part{\phi^\mu(s)}\part{\phi^\nu(t)}\hst{s}{t}.}
When $\mu=\nu$, this is the same expression we had when $\beta=0$, except that
here we have $\gmm$ instead of $G$, and here $h_{\lij,P_k}$ can be a graph
with both $x$ vertices and $y$ vertices.  When $\mu\ne\nu$, we obtain a new
contribution to the Ward identity.  Using eqn. \ddsGxy\ for $\der s\gxy(s)$,
we can write this new term as
\eqn\gammamnLdef{\eqalign{\delta_n\Gamma_C(\lij,P_k) =
                -\emn i\bab\int\dtp s&\dtp t f_n(s)
                 \left[\de(s-t)-1\right]\cr
                 &\times\part{\phi^\mu(s)}
                 \part{\phi^\nu(t)}\hst s t.}}
Thus, once we add the uniform magnetic field, there are four distinct
types of graphs for $\delta\Gamma$.  They can be represented by
\eqn\dnGdiag{\eqalign{\delta_n\Gamma=&\,
             \hbox{\lower.33in\hbox{\epsffile{Magtwoa.eps}}}\,
             +{\partial\over\partial x(s)}{\partial\over\partial x(t)}
             \,\hbox{\lower.33in\hbox{\epsffile{Magtwob.eps}}}\cr
             \noalign{\bigskip}
            -&{\partial\over\partial x(s)}{\partial\over\partial y(t)}
            i\epsilon^{xy}\bab\left[\,\,
             \hbox{\lower.33in\hbox{\epsffile{Magtwoc.eps}}}
            \,\,\right],}}
(and similarly with $x$ and $y$ interchanged.)

\subsec{The Kinetic Term}
The changes to the third term of the Ward identity are more complicated.
As we have seen, the third term of the Ward identity can be described
as inserting a $\delta_nH_0$ into 1PI diagrams and it arises from
truncating an $\alpha_{-m}$ and $\alpha_{m-n}$ from connected diagrams.
However, now we have two different ways of inserting a $\delta H$,
coming from the two terms of
\eqn\wxxwyy{{\partial^2W\over\partial\vec\alpha_{-m}\cdot
                             \partial\vec\alpha_{m-n}}
           ={\partial^2W\over\partial\alpha_{-m}^x\partial\alpha_{m-n}^x}
           +{\partial^2W\over\partial\alpha_{-m}^y\partial\alpha_{m-n}^y}.}
For the first term, we must join $\delta H_0$ to two vertices from which
$\alpha^x$'s have been truncated.  In terms of the 1PI diagrams, this
can be described as an insertion of $\delta H_x(s-s')$,
where $\delta H_x(s-s')$
is defined to be $\delta H_0(s-s')$ with $x$ vertices at $s$ and $s'$.
(This is depicted in the first diagram in equations (5.11)
and (5.21).)
Similarly, the second term is given by an insertion of
$\delta H_y(s-s')$ into 1PI diagrams, with $\delta H_y(s-s')$ defined in
the same way.

First we will calculate the effect of inserting
$\delta H_x(s'-t')+\delta H_y(s'-t')$ into a $\gmn(s-t)$ propagator when
$\mu=\nu$.  If we put $\delta H_x(s'-t')+\delta H_y(s'-t')$ between
two like vertices, say two $x$-vertices, then for a particular graph
we obtain the following contribution to the third term in the Ward identity.
\eqn\Donedef{\eqalign{D_1(\lij,P_k)=-&\int\dtp s\dtp{s'}\dtp t\dtp{t'}
                    \part{x(s)} \part{x(t)} \hst s t \cr
           &\qquad\times\left[\gxx(s-s')\delta H_x(s'-t')\gxx(t'-t)\right.\cr
           &\left.\quad\qquad+\gxy(s-s')\delta H_y (s'-t')\gyx(t'-t)\right]\cr
           =-&\part{x(s)} \part{x(t)}\left[\,\,
            \hbox{\lower.33in\hbox{\epsffile{Magthree.eps}}}\,\,
            \right].}}
To include the case when $s=t$, we define $\hst s s= {1\over2}\hs s$,
and when $s=t$ we omit the integration over $t$.  Since $\gxx$ is just a
rescaling of $G$ and $\delta H_x = \delta H_0$, we can write the first term
in the square brackets as
\eqn\Dfirsterm{{\alpha^4\over(\alpha^2+\beta^2)^2}
            G(s-s')\delta H_0(s'-t')G(t'-t).}

The $s'$ and $t'$ integrals over the second term in the square brackets
are given by
\eqn\Ixxdef{\eqalign{I_{xx}=&\int\gxy(s-s')\delta_fH_0(s'-t')
                             \gyx(t'-t)\dtp{s'}\dtp{t'}\cr
                     =&-\int\dtp{s'}\dtp{t'}\left(\der{s'}\gxy(s-s')\right)
                          f_n(s')\left[H_0(s'-t')\gyx(t'-t)\right]\cr
                     &-\int\dtp{s'}\dtp{t'}\left[\gxy(s-s')H_0(s'-t')\right]
                         f_n(t')\der{t'}\gyx(t'-t).}}
We can use the identities given in equation \ddsGxy\ and \HGxy\ to write
$I_{xx}$ as
\eqn\Ixxcalc{\eqalign{I_{xx}=
        &-{\beta^2\alpha^2\over(\alpha^2+\beta^2)^2}
        \int\dtp{s'}\dtp{t'}G(s-t')H_0(t'-s')f_n(s')\der{s'}G(s'-t)\cr
       &-{\beta^2\alpha^2\over(\alpha^2+\beta^2)^2}
        \int\dtp{s'}\dtp{t'}\left[\der{t'}G(s-t')\right]f_n(t')H_0(t'-s')
                                                 G(s'-t).}}
On comparing this with equation \fullI, we see that $I_{xx}$ is proportional
to the integral $I$ that we evaluated when $\beta=0$.  Then $I_{xx}$ is
given by
\eqn\finalIxx{\eqalign{I_{xx}=
           &{\beta^2\alpha^2\over(\alpha^2+\beta^2)^2}I(s,t)\cr
          =&{\beta^2\alpha^2\over(\alpha^2+\beta^2)^2}
           \int\dtp{s'}\dtp{t'} G(s-s')\delta_fH_0(s'-t')G(t'-t).}}
Therefore, we can write $D_1$ as
\eqn\finalDone{\eqalign{D_1(\lij,P_k)=
            -&\left[{\alpha^4\over(\alpha^2+\beta^2)^2}
            +{\beta^2\alpha^2\over(\alpha^2+\beta^2)^2}\right]\cr
           \times&\int\dtp s\dtp{s'}\dtp t\dtp{t'}\part{\phi(s)}\part{\phi(t)}
            \hst s t \cr
            &\qquad\times\left[G(s-s')\delta H_0(s'-t')G(t'-t)\right]\cr
            =-&{\alpha^2\over\alpha^2+\beta^2}\part{\phi(s)}\part{\phi(t)}
            \,\hbox{\lower.33in\hbox{\epsffile{Magfour.eps}}}.}}
This is just ${\alpha^2\over\alpha^2 + \beta^2}$ times the expression we
originally had for the third term of the Ward identity when there was no
magnetic field.  Now we can use our original calculations of $D_P$ and
$D_L$ when there was no magnetic field to find the analogous terms when
$\beta\ne0$.  They will just be rescaled by ${\alpha^2\over\alpha^2+\beta^2}$,
so they can be written as
\eqn\DPzmag{D_{P,0}(\lij,P_k)=\int\dtp s\dtp{s'}
          {\partial^2\over\partial \phi_\mu^2(s)}\hs s f_n'(s')\gmm(s-s')}
\eqn\DPdeltmag{D_{P,\Delta}(\lij,P_k)=-\aab\int\dtp s
          {\partial^2\over\partial \phi_\mu^2(s)}\hs s f_n'(s)}
\eqn\DLzmag{\eqalign{D_{L,0}(\lij,P_k)={1\over2}\int&\dtp s\dtp t\dtp{s'}
                     \part{\phi^\mu(s)}\part{\phi^\mu(t)}\hst s t\cr
                  &\times f_n'(s')\left[\gmm(s-s')+\gmm(t-s')\right]}}
and
\eqn\DLdeltmag{\eqalign{D_{L,\Delta}(\lij,P_k)
                   ={1\over2}&\int\dtp s\dtp t\dtp{s'}
                     \part{\phi^\mu(s)}\part{\phi^\mu(t)}\hst s t f_n(s')\cr
                    \times&\left[\de(t-s')\der{s'}\gmm(s-s')
                           +\de(s-s')\der{s'}\gmm(t-s')\right].}}
The diagrams for these four expressions are basically the same as the
ones given in \DPdiag\ and \DLdiag.

Next, we will calculate the contribution to the Ward identity when
$\delta H_x(s' -t')+\delta H_y(s'-t')$ is inserted into a $\gxy(s-t)$
propagator.  In that case, a particular graph has the value
\eqn\Dtwodef{\eqalign{D_2(\lij,P_k)=-&\int\dtp s\dtp t
                 \part{x(s)}\part{y(t)}\hst s tI_{xy}\cr
                 =-&\part{x(s)}\part{y(t)}\left[\,\,
                 \hbox{\lower.33in\hbox{\epsffile{Magfive.eps}}}
                 \,\,\right],}}
with $I_{xy}$ given by
\eqn\Ixydef{\eqalign{I_{xy}=\int\dtp{s'}\dtp{t'}
               &\Bigl[\gxx(s-s')\delta H_x(s'-t')\gxy(t'-t)\cr
              &\quad+\gxy(s-s')\delta H_y(s'-t') \gyy(t'-t)\Bigr].}}
We can evaluate $I_{xy}$ in the same way we calculated $I_{xx}$,
with the result that
\eqn\finalIxy{I_{xy}=0.}
Therefore, inserting $\delta H_0$ into an off-diagonal propagator, $\gmn$
with $\mu\ne\nu$, contributes nothing new to the Ward identity.

\subsec{The Zero-Mode Term}
Finally, we turn our attention to the last term in the Ward identity.
This term comes from connected diagrams with a truncated $\alpha_{-n}^\mu$
and an extra $\part{\phi^\mu}$ derivative acting on a vertex.  Because
we are working in two dimensions, there are now three possibilities for these
diagrams.  For the connected graph given by

\eqn\magcondiag{\vcenter{\epsffile{Magsix.eps}},}
they are represented by
\eqn\Cmagdiag{\eqalign{C=&\quad\vcenter{\epsffile{Magsevena.eps}}\,
               +\,\vcenter{\epsffile{Magsevenb.eps}}\cr
              \noalign{\bigskip}
          &\qquad\qquad+\qquad\vcenter{\epsffile{Magsevenc.eps}}.}}

The first two diagrams correspond to the case when the truncated
$\alpha_{-n}^\mu$ was connected to the
graph with a diagonal propagator, $\gmm$.  Then the evaluation of the
graph proceeds in exactly the same way it did when there was no magnetic
field, except that $G$ is replaced with $\gmm$.  Once again, the contribution
from these graphs cancels the ``zero mode'' contribution, $D_{P,0}+D_{L,0}$,
from the third term in the Ward identity.

In the remaining case, the truncated $\alpha_{-n}$ was connected to the graph
by an off-diagonal propagator.  For example, if the truncated $\alpha_{-n}^x$
was connected to the graph with $\gxy$, then the expression for the graph is
\eqn\Cthreedef{\eqalign{C_3(\lij,P_k)=&\int\dtp s\dtp t\dtp{s'}
                              f'(s')\gxy(s'-s)\hsta s t\cr
             =&\,\,\hbox{\lower.33in\hbox{\epsffile{Mageight.eps}}},}}
where now $\hsta s t$ is the integral over a connected graph with an extra
$\part{y_0}$ acting on the $V(y_0)$  at the $s$-vertex and an extra
$\part{x_0}$ at the $t$ vertex. For the third graph in \Cmagdiag, $\hsta s t$
is given by
\eqn\maghstadiag{\eqalign{\hsta s t =&
           \quad\vcenter{\epsffile{Magninea.eps}}\cr
         \noalign{\bigskip}
          =&\quad\vcenter{\epsffile{Magnineb.eps}}.}}
We can easily perform the $s'$ integral,
using $f_n(s') = ie^{ins'}$ and the Fourier series for $\gxy$.  It is
\eqn\fGint{\int\dtp{s'}f_n'(s')\gxy(s'-s)=i\bab e^{-\epsilon|n|}f_n(s).}
Then the contribution to the last term in the Ward identity due to
$\hsta s t$ can be written as
\eqn\finalCthree{\eqalign{C_3(\lij,P_k)=&i\bab e^{-\epsilon|n|}\int\dtp s\dtp t
                 \left[f_n(s)-f_n(t)\right]\hsta s t\cr
                =& i\bab e^{-\epsilon|n|}\left[\,\,
                 \hbox{\lower.33in\hbox{\epsffile{Magten.eps}}}
                 \,\,\right],}}
where we have also included the graph where an $\alpha_{-n}^y$ was truncated
from the vertex at $t$.

We would like $C_3$ to cancel the part of $\delta_n\Gamma_C$ that
came from treating the zero mode separately, namely, the part that comes from
the $-1$ in the expression $\de -1$ for $\partial\gmn(s)/\partial s$.
This is given by
\eqn\gammazdef{\delta_n\Gamma_Z(\lij,P_k) = -i\bab\int\dtp s\dtp t
        \left[f_n(s)-f_n(t)\right]\part{\phi^y(s)}\part{\phi^x(t)}\hst s t.}
Once again, we have the difficulty that $C$ is a function of $\alpha$
and $\Gamma$ is a function of $\phi$.  However, if we Legendre transform
$\delta_n\Gamma$, then for each $\hstat s t$ we obtain a corresponding
graph from a $\part{\phi^y(s)}\part{\phi^x(t)}\hst s t$.  Even so, the two
graphs cancel only to the leading order in $\epsilon$, because the $C_3$ has an
extra factor of $e^{-\epsilon|n|}$.  From discussions similar to those
in the following sections, we can show that
$\delta_n\Gamma_Z - C_3\rightarrow 0$
as $\epsilon\rightarrow 0$,
so that, effectively, the last term in
the Ward identity does cancel the zero-mode part of $\delta\Gamma$.

\subsec{Final Form of the Ward Identity}
As in the case when there was no magnetic field, we must now sum over all
locations of the insertion of $\delta H$ and over all graphs with $N$ vertices.
Again, we sum the $\lij$'s and $P_k$'s from zero to infinity and then subtract
off the one-particle reducible graphs.  We find that the Ward identity is
given by
\eqn\finalWId{W_P+W_L+W_C+W_Z = 0.}
$W_P$ comes from varying the petals of a graph and is given by
\eqn\WP{\eqalign{W_P=&\sum_{s_k}\sum_{\lij,P_k}\delta_n\Gamma_P(\lij,P_k)
                            -D_{P,\Delta}(\lij,P_k)\cr
                 =&\sum_{k}\siint f_n'(s_k)\left(1+\aab\dal_k\right)h(\phi).}}
$W_L$ comes from varying the diagonal propagators and has the form
\eqn\WL{\eqalign{W_L=&\sum_{s_k,s_m}\sum_{\lij,P_k}
                    \delta_n\Gamma_L(\lij,P_k) -D_{L,\Delta}(\lij,P_k)\cr
           =&\sum_{k\ne m}\siint\part{\phi^{\mu}(s_k)}\part{\phi^{\mu}(s_m)}
             h(\phi)\cr
    &\left\{f_n(s_k)\der{s_k}\gmm(s_k-s_m)-\int\dtp{s'}f_n(s')
    \de(s_k-s')\der{s'} \gmm(s_m-s') \right\}.}}
$W_C$ is due to varying the off-diagonal propagators and is given by
\eqn\WC{\eqalign{W_C=&\sum_{s_k,s_m}\sum_{\lij,P_k}
         \delta_n\Gamma_C(\lij,P_k)-\delta_n\Gamma_Z(\lij,P_k)\cr
                  =&-\emn i\bab\sum_{k,m}\siint f_n(s_k)\de(s_k-s_m)
                      \part{\phi^\mu(s_k)}\part{\phi^\nu(s_m)} h(\phi).}}
Lastly, $W_Z$ is due to the treatment of the zero-mode in the off diagonal
propagator and is given by
\eqn\WZ{\eqalign{W_Z =&\emn i\bab(1-e^{-\epsilon|n|})\cr
                &\times\sum_{s_k,s_m}
                \siint \left[f_n(s_k)-f_n(s_m)\right]\part{\phi^\mu(s_k)}
                     \part{\phi^\nu(s_m)} h(\phi).}}
In these equations, $h(\phi)$ is due to summing over all graphs with $N$
vertices. Before we subtract off the contribution due to
disconnected and one-particle reducible graphs, it has the form
\eqn\hphidef{h(\phi)=\nf\prod_{i<j}e^{\gmn(s_i-s_j)\part{\phi^\mu(s_i)}
                      \part{\phi^\nu(s_j)}}\prod_ie^{{1\over2}\gmm(0)\dal_i}
                        V(\vec \phi(s_i)).}
In summary, the Ward identity is given by a sum over the following
graphs:

\eqn\finalWIdiag{\eqalign{0=&\left(1+\aab\dal_{s_k}\right)\,
                \hbox{\lower.33in\hbox{\epsffile{Magelevena.eps}}}\cr
               \noalign{\bigskip}
              &+\part{x(s_k)}\part{x(s_m)}\left[\,\,
               \hbox{\lower.33in\hbox{\epsffile{Magelevenb.eps}}}
                \,\,\right]\cr
               \noalign{\bigskip}
              &-i\bab\epsilon^{xy}\part{x(s_k)}\part{y(s_m)}\!\!\left[\,
               \hbox{\lower.33in\hbox{\epsffile{Magelevenc.eps}}}
               \,\right]\cr
               &+\{x\leftrightarrow y\},}}

where we have ignored $W_Z$ because it is $\propto \epsilon$, and we
have obtained the last graph by interchanging $k$ and $m$ (and $x$ and
$y$) in $W_C$.

\subsec{Comments}
At this point, we would like to make a few comments about the Ward identity.
First, at order $V_0$, it takes the form
\eqn\WIVone{0=\siint f_n'(s)\left[1+\aab\dal_s\right]e^{{1\over2}\gmm(0)\dal_s}
                              V(\vec\phi(s)).}
This gives us the $O(V_0)$ tachyon equation of motion,
\eqn\Veom{\left(1+\aab\dal_s\right)V(\vec\phi(s))=0.}
This equation of motion is satisfied whenever $V$ is periodic in $x$ and
$y$ with period $2\pi/k$, where $k$ satisfies
\eqn\kcond{\aab k^2 = 1.}
This is exactly the condition for the phase transition predicted by the
simple renormalization group arguments in reference \cff.

To this order, the Ward identity is also trivially satisfied whenever
\eqn\trivdef{e^{{1\over2}\gmm(0)\dal_s}V(\vec\phi(s))=0.}
Using the definition for $\gmm(t)$ in equation \Gmmdef\ and writing
$V(\vec\phi(s))$ in terms of Fourier modes, we find that this condition is
\eqn\trivcalc{e^{{1\over2}\aab(\ln\epsilon^2)k^2}
            V_{\vec k}e^{i\vec k\cdot\vec\phi} = 0.}
In the limit as $\epsilon$ goes to zero, this becomes
\eqn\trivial{\epsilon^{\aab k^2}V_{\vec k} = 0.}
If the potential is normal ordered, then the left-hand side of eqn. \trivial\
is
non-zero, and $V$ must have the period given by eqn \kcond.  However, if
instead
we choose $V_{\vec k}\epsilon$ to be finite for any $k$, then
when $k^2 \aab > 1$  this second condition is automatically true.  This
second choice for regulating $V_{\vec k}$  occurs when we start with a
discrete lattice for time and then pass from a finite sum over time to
an integral.  For calculating $\beta$-functions, this is a natural
choice to make for $V_{\vec k}$.   Therefore, in the following section
we will define $T_{\vec k} = V_{\vec k}\epsilon$; and, whenever $k^2 \aab
>1$, we will assume that $T_{\vec k}$ is finite.

More generally, as long as $V(\vec\phi(s))$ satisfies the equation of motion
given in eqn.
\Veom, then $W_p$, the contribution to the Ward identity due to varying a
petal is always zero.
Furthermore, $W_L$ and $W_C$ would also automatically be zero if we could
immediately replace $\de$ with a delta function, $\delta$.  This is actually
possible for any individual graph.  Because all the propagators in the
graph are logarithms or sign functions, one can show by power counting
that the graph diverges at most logarithmically.  Thus, we only need to
keep the lowest term in $\epsilon$ for $\Delta_\epsilon$, so we can
replace it with a $\delta$-function.  Consequently, to any finite order
in $\alpha'$, the Ward identity is satisfied whenever $k^2
\alpha/(\alpha^2 +\beta^2) =1$.  Once we perform the sum over all
numbers of links and petals, which amounts to summing to all orders in
$\alpha'$, we obtain the graphs described by $h(\phi)$ in equation
\hphidef.  In these graphs,
the vertices are joined by the propagators
$E(t;\mu,\nu)=\exp\left[G^{\mu\nu}(t)\right]$, which go as $1/t^2$ when
$\mu=\nu$.  Naive power counting tells us that now the graphs can
diverge as $1/\epsilon$;  and in reference \cff, we found that the
free energy does, in fact, diverge as $1/\epsilon$.  Therefore, we
cannot just set $\Delta_\epsilon=\delta$, and, in the following
section, we will proceed more carefully.

\def \lij{l_{ij}}
\def \siint{\int \prod_i {ds_i\over 2\pi}}
\def \riint{\int \prod_i {dr_i\over 2\pi}}
\def \symf{{1\over \prod_{i,j}\lij!\prod_i P_i!2^{P_i}}}
\def \nf{{1\over N!}}
\def \gpi{\left[ G(0)\right]^{\sum_i P_i}}
\def \glij{\prod_{i,j}\left[G(s_i-s_j)\right]^{\lij}}
\def \part#1{{\partial \over \partial{#1}}}
\def \dtp#1{{d{#1}\over 2\pi}}
\def \hs#1{h_{\lij,P_k}({#1})}
\def \hsa#1{\tilde h_{\lij,P_k}({#1},\alpha)}
\def \hsat#1{\tilde h_{\tilde \lij,\tilde P_k}({#1},\alpha)}
\def \hst#1#2{h_{\lij,P_k}({#1,#2})}
\def \hsta#1#2{\tilde h_{\lij,P_k}({#1,#2,\alpha})}
\def \hstat#1#2{\tilde h_{\tilde \lij,\tilde P_k}({#1,#2,\alpha})}
\def \de{\Delta_\epsilon}
\def \detwo{\Delta_{2\epsilon}}
\def \gmn{G^{\mu\nu}}
\def \gmm{G^{\mu\mu}}
\def \gmme{\gmm_{2\epsilon}}
\def \gxx{G_{xx}}
\def \gyy{G_{yy}}
\def \gxy{G_{xy}}
\def \gyx{G_{yx}}
\def \emn{\epsilon^{\mu\nu}}
\def \aab{{\alpha\over\alpha^2 + \beta^2}}
\def \bab{{\beta\over\alpha^2 + \beta^2}}
\def \der#1{{d\over d{#1}}}
\def \vk{\vec k}
\def \vp{\vec\phi}
\def \sign{{\rm sign}}

\newsec{The Flows Generated by the Ward Identity}
In the following sections, we will show that, to leading order in
$\epsilon$, the condition for the Ward identity to be satisfied is the
same as the condition for the $\beta$-function, given by
$\epsilon \part\epsilon\Gamma$,
to equal zero.  Also, we will prove that the $\cos x$ potential satisfies the
Ward identity at every order in $V_0$ and to all orders in $\alpha'$
whenever $\aab=1$ and $\bab\in Z$.

To show the equivalence of the Ward identity to the equation
$\epsilon\part\epsilon\Gamma=0$ for any arbitrary potential, $V(\vec\phi(t))$,
we begin by writing $V(\vec\phi(t))$ in terms of its Fourier modes,
$V(\vec\phi(t))=\sum_k (T_{\vec k}/\epsilon)
e^{i\vec k\cdot\vec\phi(t)}$.  (For simplicity, we use a sum over $k$, here,
but the results should not change if we take an integral instead.)
At order ${V_0}^N$, the Ward identity for the specific modes of $V$
labeled by $\vec k_1,\dots, \vec k_N$ is then given by
\eqn\fourierWI{W_P+W_L+W_C+W_Z = 0,}
where
\eqnn\WPk
\eqnn\WZk
\eqnn\WLk
\eqnn\WCk
$$\eqalignno{W_P=&\sum_{l}\siint f_n'(s_l)\left(1-\aab|\vec k_l|^2\right)H,
    &\WPk\cr
  W_Z =&\emn i\bab(e^{-\epsilon|n|}-1)\sum_{l<m}
          \siint \left[f_n(s_l)-f_n(s_m)\right]k_l^\mu k_m^\nu H,\qquad&\WZk\cr
  W_L=&-\sum_{l< m}\siint k_l^\mu k_m^\mu H \cr
    &\qquad\qquad\times\Biggl\{\sum_{a,b=(l,m)\atop(m,l)}
    \Bigl[f_n(s_a)\der{s_a}\gmm(s_a-s_b)&\WLk\cr
     &\qquad\qquad\qquad\qquad -\int\dtp{s'}f_n(s')\de(s_a-s')\der{s'}
    \gmm(s_b-s') \Bigr]\Biggr\},\cr
\noalign{\noindent and}
     W_C=&\emn i\bab\sum_{l,m}\siint f_n(s_l)\de(s_l-s_m)
                      k_l^\mu k_m^\nu H.&\WCk}$$
Here, $H$ is from the 1PI graphs with vertex factors
${T_j\over\epsilon}e^{i\vec k_j\cdot\vec\phi(t)}$, for $j=1,\dots, N$.
It is given by the expression
\eqn\htildedef{\tilde h=\nf
                \prod_j{T_j\over\epsilon}e^{-{1\over2}\gmm(0)k_j^2}
                 e^{i\vec k_j\cdot\vec\phi(s_j)}
               \prod_{i<j}e^{-\gmn(s_i-s_j)k_i^\mu k_j^\nu},}
minus all the disconnected and one-particle reducible graphs.  (For $W_L$
and $W_C$, we must be
a little more careful in specifying the graphs we subtract from $\tilde h$;
we want to subtract all graphs the are one-particle reducible when we
include the $\der{s_a}\gmm(s_a-s_b)$ propagator for $W_L$ and the
$\de(s_l-s_m)$ propagator for $W_C$.)  We will define the vector $\vec q$
to be $\vec q = \sum_{i=1}^N \vec k_i$.  Then the total charge of the graph
is given by $q = |\vec q|$.

For the remainder of this paper, we will find it convenient to
redefine $G^{\mu\mu}$ to be
\eqn\Gmmsdef{G^{\mu\mu}(t)= - \aab \ln\left(e^\epsilon+e^{-\epsilon}-
                               2\cos t\right).}
This is obtained from the original definition of $G^{\mu\mu}$ in equation
\Gmmdef\ by subtracting $\aab\epsilon$.  When we use this new definition of
$G^{\mu\mu}$, we must also include a factor of
$\exp\left(-{1\over2}\aab \epsilon({q_x}^2+{q_y}^2)\right)$ in
$\tilde h$.  For brevity, we will define
\eqn\Tdef{T=\exp\left(-{\epsilon\over2}\aab ({q_x}^2+{q_y}^2)\right)
         \prod_j{T_j\over\epsilon}\exp\left(-{1\over2}\gmm(0)k_j^2\right).}

To evaluate the integrals in this expression for the Ward identity, we must
Taylor expand the $e^{i\vk_j\cdot\vp(s_j)}$'s around a common point, $t$.
Thus, we have
\eqn\gexp{\prod_j e^{i\vk_j\cdot\vp(s_j)}= e^{i\vec q\cdot\vp(t)}
          + \sum_j(s_j-t)\vk_j\cdot {\partial\vp(t)\over\partial t}
            e^{i\vec q\cdot\vp(t)} + ...}
 From this expansion, we will obtain flows for coupling constants of
the operators $e^{i\vk\cdot\vp(t)}$,
$\partial\vp(t)\over\partial t$,
${\partial\vp(t)\over\partial t}e^{i\vec q\cdot\vp(t)}$,
and the higher degree operators.  These flows will be the coefficients
of these operators in the Ward identity after we
have expanded all the vertices around a common point.
In the following sections, we will show that these
flows for the relevant and marginal operators are the same as the
renormalization group flows generated by the $\beta$-function, given by
$\epsilon \partial\Gamma/\partial\epsilon$.
The lowest degree operators obtained from charge-$0$
graphs are constants and $\partial\vp\over\partial t$.  All other
operators from charge-$0$ graphs have additional derivatives, so they are
of higher degree and should be irrelevant.  Similarly, for any
charge-$\vec q$ graph with $0<|q|\le1$, the lowest order operator generated
from the Taylor series is relevant and is given by $e^{i\vec q\cdot\vp}$.
All higher order operators have at least one more derivative, so they
are irrelevant.  Lastly, we expect graphs with charge $q$ greater than one
to only generate flows for irrelevant operators.

We note that from the expansion in equation \gexp, there is no way to
generate the non-local
friction term or the constant gauge field \fisher.  As a result, we do not
obtain
any flows for $\alpha$ and $\beta$.  In addition, we will show that
the flow for $\partial\vp/\partial t$ also vanishes due to symmetries of
$H$.

\newsec{Calculating the Flow for $e^{i\vec q\cdot \vec\phi}$}
$W_P.\qquad$
We will first evaluate the $W_P$ term.  If the potential satisfies the
one-loop equation of motion \kcond, then $W_P$ is automatically $0$.
For arbitrary potentials, we make the following change of
variables:
\eqn\chvarWP{\eqalign{s_+&=s_l\cr
             r_j&=s_j-s_l \qquad j=1,\dots N.}}
Then we Taylor expand around $s_+$.  To lowest order in $r_j$ and $\epsilon$,
we obtain
\eqn\finalWP{\eqalign{W_P=T\sum_l\left(1-\aab|\vk_l|^2\right)
                   \times\int\dtp{s_+}f_n'(s_+)e^{i\sum_j\vk_j\cdot\vp(s_+)}
                       F.}}
$F$ is essentially the free energy and is defined by
\eqn\Fdef{F=\int\prod_{i=1}^N \dtp{r_i}(I-R),}
where $I$ is the integrand for the partition function, given by
\eqn\Ikdef{I= \prod_{i<j}e^{-\gmn(r_i-r_j)k_i^\mu k_j^\nu};}
and $R$ is the integrand for all connected, one-particle reducible graphs
and all disconnected graphs.

$W_Z.\qquad$
Similarly, we can expand the vertices in $W_Z$ about a common point, $s_+$.
The leading term is
\eqn\expWZ{\eqalign{W_Z=-\epsilon|n|i&\bab T
               \int \dtp{s_+}f_n'(s_+)e^{i\sum_j\vk_j\cdot\vp(s_+)}\cr
       &\times\sum_{l<m}\left[k_l^\mu k_m^\nu\emn\siint(s_m-s_l)F\right].}}
We will replace $|n|if_n'(s_+)$ by $\pm f_n''(s_+)$ and integrate by parts.
Then, for $n>0$, the $s_+$ integral becomes
\eqn\spint{\int \dtp{s_+}f_n'(s_+){d\over d s_+}
                    e^{i\sum_j\vk_j\cdot\vp(s_+)}.}
Thus, the first non-vanishing contribution to $W_Z$ is for the flow of
$\der{s_+} e^{i\vec q\cdot\vp(s_+)}$ instead of the flow of
$e^{i\vec q\cdot\vp(s_+)}$.  Note that $W_Z=0$ when $\vec q=0$.  For all
other values of $q$, we obtain the flow for irrelevant operators and
therefore still expect $W_Z$ to vanish.

$W_L.\qquad$
$W_L$ is more complicated to evaluate.  We have already noted that when
$\epsilon\rightarrow0$, $\de(s_a-s')$ becomes a delta function.  As a result,
$W_L$ would be zero if the remaining part of the integrand weren't
singular.  Therefore, the non-zero part of $W_L$ should really come from
the regions of integration where the vertices are close together and the
integrand is singular, which is a justification for Taylor
expanding the vertices about a common point.  In addition, because
$\de(s_a-s')$ approaches a delta function for small $\epsilon$, we
can expand it around small $\epsilon$ and small $s_-=s_a-s'$,
(since the region of integration only covers one period of $\de$).

First, we will make the following change of variables for the second term in
the curly brackets in eqn. \WLk\ for $W_L$.
\eqn\chanvarWL{\eqalign{s_-=&s'-s_a\cr
                        s_+=&s_a\cr
                        t=&s_b-s_a\cr
                        r_j=&-s_j+s_l \qquad {\rm when}\quad (a,b)=(l,m)\cr
        {\rm and} \quad r_j=&s_j-s_l  \qquad {\rm when}\quad (a,b)=(m,l).}}
 From the definitions of $r_j$ and $t$, note that
\eqn\trl{t=-r_m \qquad {\rm and} \qquad r_l = 0.}
The change of variables for the first term in the curly brackets is
the same, except we do not need an $s_-$.
We will also define
\eqn\gdef{g_j(s)=e^{i\vk_j\cdot\vp(s)}.}
Then $W_L$ is given by
\eqn\WLchvar{\eqalign{W_L=&-T\sum_{l<m}k_l^\mu k_m^\nu\int\dtp{s_+}\dtp{s_-}
                 \prod_{j=1\atop j\ne l}^N\dtp{r_j} \cr
                 &\times\Biggl\{F(-r_1,\dots,-r_N)
                 \prod_{i=1}^Ng_i(s_+-r_i)
                 +F(r_1,\dots, r_N)
                  \prod_{i=1}^Ng_i(s_++r_i+t)\Biggr\}\cr
             &\left\{-f_n(s_+)\der t\gmm(t)\delta(s_-)
              -f_n(s_++s_-)\de(s_-)\der{s_-}\gmm(t-s_-)\right\},}}
where $F(r_1,\dots, r_M)$ is
the 1PI graphs with vertices at $r_1, \dots, r_N$ as defined in equation
\Fdef.

First we concentrate on the $s_-$ integral.  $\de(s_-)$ is acting as
an approximate delta function, so for a smooth enough function like
$f_n$, we only need to keep the first few terms in the
Taylor expansion for $s_-$.  After expanding to first order in $s_-$,
the integral of interest is
\eqn\sminusint{-\int\dtp{s_-}\left[f_n(s_+)\de(s_-)\der{s_-}\gmm(t-s_-)
              +f_n'(s_+)s_-\de(s_-)\der{s_-}\gmm(t-s_-)\right].}
The first term can be integrated exactly.  It is
\eqn\deltGint{\eqalign{\int\dtp{s_-}\de(s_-)\der{s_-}\gmm(t-s_-)=&
                     -i\sum_{m\ne0}\aab\sign(m)e^{imt}e^{-2|m|\epsilon}\cr
                   =&-\der t \gmme(t)}}
where $\gmm_{m\epsilon}(t)$ is defined to be equal to $\gmm(t)$ with
$\epsilon$ replaced by $m\epsilon$.

To integrate the second term, we must evaluate the integral for small
values of the argument of $\de(s_-)$.  In addition, we only need to consider
small values of $t$, so that the argument of $\gmm(t-s_-)$ is small.
Otherwise, if $t$ were large, $\der {s_-}\gmm(t-s_-)$ would be a slowly
varying function of $s_-$ and integrating it against $s_-$ times an
approximate delta function would just give zero.
For small $s_-$, $\epsilon$, and $t$, the functions $\de$ and
$\der{s_-}\gmm$ are given by equations \smalldelta\ and \smallG\ with
$\alpha'=\aab$ and the integral is
\eqn\sdeltGint{\eqalign{\int\dtp{s_-}s_-\de(s_-)\der{s_-}\gmm(t-s_-)
            \approx&-\epsilon\aab{4\epsilon\over(2\epsilon)^2+t^2}\cr
            =&-\aab\epsilon\detwo(t)\left(1+O(\epsilon^2,t^2)\right).}}
Substituting these integrals back into the expression for $W_L$, using
equation \trl, and making
the change of variables $r_i\rightarrow-r_i$, $t\rightarrow-t$ in the
first term in the first curly brackets, we obtain
\eqn\WLmess{\eqalign{W_L=&-T\sum_{l<m}k_l^\mu k_m^\mu\int\dtp{s_+}\dtp t
       \prod_{j=1\atop j\ne m,l}^N\dtp{r_j}
        F(r_1,\dots, r_N)\cr
   &\left\{\left[\prod_{i=1\atop i\ne l,m}^N g_i(s_++r_i+t)g_m(s_+)g_l(s_++t)
    -\!\!\prod_{i=1\atop i\ne l,m}^N\! g_i(s_++r_i)g_m(s_+-t)g_l(s_+)\right]
      \right.\cr
  &\quad\times\left[-f_n(s_+)\der t\gmm(t) + f_n(s_+)\der t\gmme(t)\right]\cr
    &+\left[\prod_{i=1\atop i\ne l,m}^N g_i(s_++r_i)g_m(s_+-t)g_l(s_+)
  +\!\!\prod_{i=1\atop i\ne l,m}^N\! g_i(s_++r_i+t)g_m(s_+)g_l(s_++t)\right]\cr
  &\quad\left.\times\aab f_n'(s_+)\epsilon\detwo(t)
     \vphantom{\prod_{i=1\atop i\ne l,m}^N}\right\},}}
where the second term in the curly brackets is one order higher in $\epsilon$
than the first term.  Now we want to Taylor expand the $g_i$'s around
$t=0$ and $r_i=0$.  For the first term in the curly brackets, we keep the
linear
order terms and for the second term in the curly brackets we keep only the
lowest order, constant terms, because it is already one order higher in
$\epsilon$.  Concentrating on the first term in the curly brackets, we see
that the zeroth order terms from the expansion in $t$ and $r$ all cancel.
Similarly, the first order terms in $r_i$ also all cancel, so we are left
only  with the first order terms in $t$, which are
\eqn\gder{\prod_{i=1\atop i\ne l,m}^Ng_i(s_+)g_m'(s_+)g_l(s_+)t
          +\sum_{j=1\atop j\ne m}^N\prod_{i\ne j}g_i(s_+)g_j'(s_+) t
          = t \der{s_+}\prod_{i=1}^N g_i(s_+).}
Then, substituting this back into the expression for $W_L$ and integrating by
parts, we obtain
\eqn\WLpenult{W_L= -T\sum_{l<m}k_l^\mu k_m^\mu\int\dtp{s_+} f_n'(s_+)
                         \prod_{i=1}^N g_i(s_+)
                     \int\dtp t\prod_{j=1\atop j\ne m,l}\dtp{r_j}
                    F(r_1,\dots, r_N)\times E,}
where $E$ is given by the expression
\eqn\Edef{E=t\der t\gmm(t)-t\der t\gmme(t)+2\aab\epsilon\detwo(t).}
For large values of $t$, this expression is well defined and goes to
$0$ as $\epsilon\rightarrow0$.  Therefore, as long as the rest of the
integrand is not too singular as $\epsilon\rightarrow 0$, we can
evaluate $E$ for small values of $t$ and $\epsilon$.  Then $E$ is
given by
\eqn\Eeval{\eqalign{E=&2\aab
         \left[-{t^2\over t^2+\epsilon^2}+{t^2\over t^2+(2\epsilon)^2}
                +{4\epsilon^2\over(2\epsilon)^2+t^2}\right]
              \left[1+O(t^2)+O(\epsilon^2)\right]\cr
             =&2\aab{\epsilon^2\over t^2+\epsilon^2}
              \left[1+O(t^2)+O(\epsilon^2)\right].}}
It will be useful to express this in terms of a function defined on a
circle, so, making use of the fact that it contributes to the integral
only for small $t$, we will write it as
\eqn\Efinal{\eqalign{E\approx& \epsilon\aab
      {e^\epsilon-e^{-\epsilon}\over e^\epsilon+e^{-\epsilon}-2\cos t}\cr
                =&-\epsilon\der\epsilon \gmm(t)\left[1+O(\epsilon^2)+
                   O(t^2)\right].}}
Then we can write $W_L$ in its final form as
\eqn\finalWL{\eqalign{W_L=T&\sum_{l,m}k_l^\mu k_m^\mu\int\dtp{s_+}
                f_N'(s_+)\prod_{i=1}^Ng_i(s_+)\cr
           &\times\int\prod_{j=1}^N\dtp{r_j}
         F(r_1,\dots, r_N)
        \left(\epsilon\der\epsilon\gmm(r_l-r_m)\right),}}
where we have used translational invariance to introduce $r_l$ as a dummy
variable and used the definition in equation \trl\ for $t$.

$W_C.\qquad$
All that remains now is to evaluate $W_C$.  Once again, we perform a
change of variables given by
\eqn\chanvarWC{\eqalign{s_+=&{1\over2}(s_l+s_m)\cr
                  t=&s_m-s_l\cr
                  r_j=&s_j-s_l.}}
Then $W_C$ can be written as
\eqn\WCmess{\eqalign{W_C=i&\bab T\sum_{(l,m)}k_l^x k_m^y\int\dtp{s_+}
                \prod_{i\ne l,m}\dtp{r_i}
                \prod_{j=1}^Ng_j(s_++r_j-{1\over2}t)\cr
              \times&\left[f_n(s_+-{1\over2}t)-f_n(s_++{1\over2}t)\right]
              \de(t)F(r_1,\dots, r_N).}}
When this expression is expanded around $t=0$ and $r_j=0$ to first order,
only one term remains, and $W_C$ becomes
\eqn\WCcalc{\eqalign{W_C=-i&\bab T\sum_{(l,m)}k_l^x k_m^y\int\dtp{s_+}
         f_n'(s_+)\prod_{j=1}^Ng_j(s_+)\cr
         \times&\int\prod_{i=1}^N\dtp{r_i}
              F(r_1,\dots, r_N)
         (r_m-r_l)\de(r_m-r_l).}}
Now we would like to write the following expression in a slightly different
form.
\eqn\Edeftwo{\eqalign{E=&i\bab(r_m-r_l)\de(r_m-r_l)\cr

=&-i\bab(r_l-r_m){\sinh\epsilon\over\cosh\epsilon-\cos(r_l-r_m)}.}}
For small $\epsilon$, $\de$ picks out only small values of its argument,
$r_l-r_m$, in the integral.  Therefore, to the order in $\epsilon$
that we have been doing our calculation, we can replace
$(r_l-r_m)\sinh\epsilon$
with $\epsilon\sin(r_l-r_m)$.  Then $E$ is
\eqn\finalEtwo{\eqalign{E\approx&-\epsilon i\bab
                       {\sin(r_l-r_m)\over\cosh\epsilon-\cos(r_l-r_m)}\cr
                    =&-\epsilon\der\epsilon\gxy(r_l-r_m).}}
Substituting the expression for $E$ back into the equation for $W_C$, we
obtain the final form for $W_C$.
\eqn\finalWC{\eqalign{W_C=&T\sum_{l<m}k_l^\mu k_m^\nu\int\dtp{s_+}
                 f_n'(s_+)\prod_{j=1}^Ng_j(s_+)\cr
                &\times\int\prod_{i=1}^N\dtp{r_i}
                 F(r_1,\dots, r_N)
                 \left[\epsilon\der\epsilon\gmn(r_l-r_m)\right].}}

\newsec{The Ward Identity and the $\beta$-function}
To leading order in $\epsilon$, we have obtained the following form
for the Ward identity.
\eqn\penWI{\eqalign{0=&\int\dtp{s_+}f_n'(s_+)e^{i\sum_j\vk_j\cdot\vp(s_+)}
                         e^{-{1\over2}\sum_j|\vk_j|^2\gmm(0)}
                  \exp\left(-{\epsilon\over2}\aab({q_x}^2+{q_y}^2\right)\cr
               &\prod_j{T_j\over\epsilon}\riint\Biggl\{
                 \left[I(r_1,\dots, r_N)-R(r_1,\dots, r_N)\right]\cr
               &\times\left[\sum_l\left(1-\aab\vk_l\cdot\vk_l\right)
             +\epsilon\sum_{l<m}k_l^\mu k_m^\nu\der\epsilon\gmn(r_l-r_m)\right]
               \Biggr\},}}
where $I$ is given by
\eqn\Imndef{I(r_1,\dots r_N)=\prod_{i<j}e^{-\gmn(s_i-s_j)k_i^\mu k_j^\nu}.}
$R$ is the corresponding integrand for all the connected one-particle
reducible and disconnected diagrams, with the extra condition
that when it is multiplied by $\der\epsilon\gmn(r_l-r_m)$, this
additional propagator is included in determining whether or not a
graph is 1PI or connected.

At this point  we have calculated the ``flow'' for
$\prod_{j=1}^Ng_j(s)=e^{i\sum_j\vec k_j\cdot\vec\phi(s)}$.  It is given
by the coefficient of $\int(ds_+/2\pi)f_n'(s_+)\prod_{j=1}^Ng_j(s_+)$ in
equation \penWI.  As we shall show next, to leading order in $\epsilon$,
this agrees exactly with the flows
produced by the $\beta$-function when $\sum_j \vec k_j\ne0$.
We begin by showing that the expression
for the Ward identity in equation \penWI\ is a total
derivative.  From the definition of $I$ in equation \Imndef, we can see that
\eqn\deIR{\eqalign{\left[I(r_1,\dots,r_N)-R(r_1,\dots, r_N)\right]&
         \epsilon\sum_{l<m}k_l^\mu k_m^\nu\der\epsilon\gmn(r_l-r_m)\cr
          =&-\epsilon\der\epsilon(I-R).}}
To show that the other term in the Ward identity is also a derivative
with respect to $\epsilon$, we use the definition
\eqn\Gmmzero{\gmm(0)=-\aab\ln(e^{\epsilon}+e^{-\epsilon}-2)
                    \approx -\aab\ln\epsilon^2}
to evaluate
\eqn\deGzero{\eqalign{&e^{-{1\over2}\sum_j\vk_j\cdot\vk_j\gmm(0)}
        \prod_j{T_j\over\epsilon}\sum_l\left(1-\aab\vk_l\cdot\vk_l\right)\cr
        =&-\epsilon\der\epsilon
        \left[e^{-{1\over2}\sum_j\vk_j\cdot\vk_j\gmm(0)}
       \prod_j{T_j\over\epsilon}\right]\left(1+O(\epsilon^2)\right).}}
In this calculation, we have neglected the factor of
$\exp\left(-{\epsilon\over2}\aab({q_x}^2+{q_y}^2\right)$ in equation \penWI.
However, it only contributes corrections that are proportional
to $\vec q \cdot \vec q \epsilon$, which means that, to the order
in $\epsilon$ that we are calculating, we can ignore this factor.

Thus, the Ward identity can be expressed as a total derivative
with respect to $\epsilon$.  It is
\eqn\finalWI{0=-\epsilon\der\epsilon\int\der{s_+}f_n'(s_+)
                        e^{i\sum_j\vk_j\cdot\vp(s_+)}
            \prod_j{T_j\over\epsilon}e^{-{1\over2}\sum_j|\vk_j|^2\gmm(0)}
               \riint F.}
 From this equation, we see that the Ward identity is satisfied whenever
\eqn\condone{\int\dtp{s_+}f_n'(s_+)e^{i(\sum_j\vk_j)\cdot\vp(s_+)}=0}
or
\eqn\condtwo{-\epsilon\der\epsilon\prod_j{T_j\over\epsilon}
             e^{-{1\over2}\sum_j|\vk_j|^2\gmm(0)}\riint F=0\qquad
                {\rm as} \quad\epsilon\rightarrow0.}
When the total charge is zero, $\sum_j\vk_j=0$. Therefore, the first condition
becomes
\eqn\condzc{\int_{-\pi}^\pi\dtp{s_+}f_n'(s_+)=0.}
This is always satisfied because it is the integral of a derivative of a
periodic function.  When the total charge is not zero, the first condition
is not satisfied.  In that case, the Ward identity is satisfied if and
only if the second condition, given by eqn. \condtwo, is true.

The first conclusion to be drawn from this result is that this
condition for the Ward identity to be satisfied is
the same as the condition for the system to be at a zero of the beta function,
where the beta function is given by $-\epsilon\partial\Gamma/\partial\epsilon$.
 From equations \htildedef, \Fdef\ and \Ikdef, the 1-particle irreducible
function for graphs with vertices $(T_j/\epsilon) e^{\vk_j\cdot\vp(s_j)}$
for $1\le j\le N$ is given by
\eqn\onePI{\Gamma(\vp)=\prod_i{T_j\over\epsilon}
                      e^{-{1\over2}\sum_j|\vk_j|^2\gmm(0)}
              \int\prod_{i=1}^N{ds_i\over 2\pi}
              e^{i\sum_j\vk_j\cdot\vp(s_j)}F.}
If we take a derivative with respect to $\epsilon$,
and then expand all the $\vp(s_j)$'s about a common
point, we obtain
\eqn\betafcn{-\epsilon{\partial\Gamma\over\partial\epsilon}=0}
if and only if
\eqn\bfcond{-\epsilon\der\epsilon\int\dtp{s_+}
              e^{i\sum_j\vk_j\cdot\vp(s_+)}
             \prod_j{T_j\over\epsilon}e^{-{1\over2}\sum_j|\vk_j|^2\gmm(0)}
             \riint F=0.}
Since $\int\dtp{s_+} e^{i\sum_j\vk_j\cdot\vp(s_+)}\ne0$ for arbitrary
$\vp(s_+)$, we find that the beta function equals zero only when equation
\condtwo\ is satisfied.  Therefore, when $q\ne0$, the conditions for the
Ward identity to be satisfied and for the system to lie at a zero of the
beta function are the same, at least in the limit as $\epsilon\rightarrow0$.

Secondly, we note that the Ward identity was an infinite set of symmetries;
for each $f_n(s)=ie^{ins}$ there is a separate identity.  However, we have
reduced each of these equations down to one single condition, given by
equation \condtwo.  As a result, if the system possess just one of these
symmetries (or if it is at a zero of the $\beta$-function) it will possess all
of the symmetries, as in the case of scale invariance leading to
conformal invariance in 2D theories.
\newsec{Validity of Approximations}
At this point, we will review the approximations that were made in obtaining
equation \finalWI\ and show that they should be valid as $\epsilon$ goes to
zero.  The first comes from expanding the vertices around a
common point, $s_+$.  This approximation involved keeping only the lowest
order terms in the Taylor series for $\prod_j g_j(s_++r_j)$ and
$f_n'(s_++t)$.  The higher order terms include higher derivatives on
$g_j$ and $f_n$.  We can always integrate by parts so that all these extra
derivatives act on the $g_j$'s.  That means we are calculationg the flows
for higher dimension operators, such as
${d\vp\over dt}e^{i\vec q\cdot\vp(t)}$,
${d^2\vp\over dt^2}e^{i\vec q\cdot\vp(t)}$, {\it etc.}  As long as $q > 0$,
all of these operators are irrelevant, so we expect them to go to zero as
$\epsilon\to 0$.  The only remaining relevant operator is ${d\vp\over dt}$,
which we will calculate in the following section.

The second kind of approximation comes from expanding propagators around
small values of their arguments and small $\epsilon$.  Before we evaluate
the integrals, all these approximations give corrections that are
$O(\epsilon^2)$ or $O(t^2)$.  The $O(\epsilon^2)$ corrections are higher
order in $\epsilon$, as claimed, but, when calculating $W_L$ and $W_C$,
we must be careful about the $O(t^2)$ corrections.  (For $W_P$, we can
directly evaluate the $O(s_-^2)$ corrections to the integral in equation
\Dpdelint, with the result that they go as at most as $O(\epsilon)$.)
Whenever we obtained a correction that was $O(t^2)$, it came
from expanding a product of an approximate delta function times a
propagator.  If only the short-distance behavior of the whole graph is
important, then this correction goes as $O(\epsilon^2)$ after we integrate
over the graph.  However, if the large distance contribution is bigger than
the short distance contribution, then the correction goes at most as
$\epsilon c$, where the $\epsilon$ comes from the approximate
$\delta$-function and the $c$ is a finite number because for large
separations of the vertices the integrals are bounded.

The third approximation involved ignoring the factor of
$\exp(-{\epsilon\over2}
\aab ({q_x}^2+{q_y}^2))$ in equation \penWI.  It gives corrections that are
$O(\epsilon)$.

To see whether these corrections should vanish as $\epsilon$ goes to zero,
we will look at the superficial degree of divergence of the graphs.  Recall
that $\exp(-k_i k_j G^{\mu\mu}(t)) \sim (\epsilon^2 +t^2)^{k_i k_j}$
for small $\epsilon$ and $t$, and $|\exp(-k_i k_j G^{\mu\nu})| = 1$ for
$\mu \ne \nu$.  For convenience, we will let $\gamma = \aab$.
Then the vertex factors and self-interactions in equation
\condtwo\ go as
\eqn\vertfac{A=\prod_{i=1}^N(T_i/\epsilon) \epsilon^{\gamma |\vk_i|^2}.}
The short-distance contributions from the integrals go as
\eqn\sddiv{S=\epsilon^{N-1+\gamma\left[q^2-\sum|\vk_i|^2\right]};}
and the large-separation contributions goes as
\eqn\lsdiv{L=\epsilon c \qquad {\rm and}\qquad L= c_p,}
for $W_L+W_C$ and $W_P$, respectively.  From equation \finalWP,
we note that when all the $\gamma|\vk_i|^2=1$, then $c_p=0$.  Also, as noted
in the previous paragraphs, the correction to $L$ should go as
$\epsilon(c+c_p)$.

If we renormalize the potential by keeping $V_i^R=T_i$
finite, then the graphs go as
\eqn\VRVI{\prod_{i=1}^N V_i^R\left[\epsilon^{\gamma q^2-1}
         +(c\epsilon+c_p)\epsilon^{\sum_{i=1}^N\left(\gamma|\vk_i^2|-1\right)}
         \right].}
Ideally, for calculating the renormalization group flow, we would like only
the short-distance behavior to be important.  In that case, the graph goes
as $\epsilon^{\gamma q^2-1}$  and the corrections go at most as
$\epsilon^{\gamma q^2}$.  Therefore, for $q\ne0$, they vanish as
$\epsilon\to0$.  Unfortunately, for arbitrary potentials, the large-distance
part of the graph can be non-negligible.  As long as $\gamma|\vk_i|^2\ge1$,
the long distance part does go to $0$ as $\epsilon\to0$, so the graphs
should be proportional to $\epsilon^{\gamma q^2-1}$ and our approximations
should be valid.  In some special cases, even when $\gamma|\vk_i|^2<1$,
the large-distance part of the graph can be renormalized away \stamp\
so that $c=c_p=0$ and the graph is still proportional to
$V^N\epsilon^{\gamma q^2-1}$.  However, in the general case, we will normal
order any vertex that has $\gamma|\vk_i^2|<1$.  In that case, $V_i^R=
T_i\epsilon^{\gamma|\vk_i|^2-1}$ is kept finite.  This is similar to what is
done in the Sine-Gordon model \amit.  Then the graph goes as
\eqn\VRnorm{\prod_{i=1}^N V_i^R\left[\epsilon^{\gamma q^2-1}
          \epsilon^{\sum_{i=1}^M\left(1-\gamma|\vk_i^2|\right)}
         +(c\epsilon+c_p)
          \epsilon^{\sum_{i=M+1}^N\left(\gamma|\vk_i^2|-1\right)}
         \right],}
where the vertices with $1\le i\le M$ have $\gamma|\vk_i|^2 <1$ and those
with $M+1\le i\le N$ have $\gamma|\vk_i|^2\ge1$.  The short-distance
contribution is now at most $\epsilon^{\gamma q^2-1}$, which means that for
$q\ne0$ the corrections to the short-distance behavior still go to zero as
$\epsilon\to0$.  The long distance behavior can still be finite, but the
corrections now go at most as $\epsilon(c+c_p)$, so they also vanish as
$\epsilon\to0$.  These estimates ignore the possibility that we could get
even more divergent contributions from the regions of integration where only
some of the variables are close.  We are assuming that once the disconnected
diagrams are subtracted, this is not the case.  Given this assumption,
we conclude that when $q\ne0$, all our approximations should be valid and
the Ward identity is equivalent to the $\beta$-function as $\epsilon\to0$.

\newsec{Charge-0 graphs and the flow for $d\vp/dt$}
The case when $\vec q=0$ is special.  The leading order contribution gives the
flow for $e^{i\vec q\cdot \vp(t)}=1$.  Because it is multiplied by equation
\condone, the Ward identity
for this operator is always equal to zero, even when we include
the corrections due to expanding the propagators and $\delta$-functions.
The next most relevant operator obtained from these graphs is $d\vp(t)/dt$.
If we repeat the calculations of the Ward identity and $\beta$-functions of
the previous sections, keeping the next higher order term in the Taylor
expansion of $f_n(s)\prod_{i=1}^N g_i(s_i)$, we find that the flow for
$\vec k_j\cdot d\vp(t)/dt$ is
\eqn\dphiflow{\riint \Biggl\{r_j
                 F(r_1,\dots, r_N)
               \times\left[\sum_l\left(1-\aab\vk_l\cdot\vk_l\right)
             +\epsilon\sum_{l<m}k_l^\mu k_m^\nu\der\epsilon\gmn(r_l-r_m)\right]
               \Biggr\},}
and all our approximations should be valid.

For each graph that contributes to this integral, there is another graph
with the signs of all the $k_i$ reversed.  Both graphs have the same value
of $F(r_1,\dots,r_N)$, and when the total charge of the graph is $0$, both
graphs contribute to the flow of $d\vp/dt$.  Therefore, when we add the
expressions from equation \dphiflow\ due to these two graphs, we obtain
zero.  We conclude that the coupling constant for $d\vp/dt$ does not flow.

\newsec{Critical cosine potential}
Finally, we return to the special case of the on-shell cosine potential
which satisfies $\aab|\vec k_i|^2 = 1$.  From the discussion in the previous
sections, we expect that the charge-one graphs for $\Gamma$ should go as
$\epsilon^0$.  This means they can have a logarithmic divergence in
$\epsilon$.  It follows that when we take $\epsilon
\partial\Gamma/\partial\epsilon$ in order to evaluate the $\beta$-function,
we can obtain a non-zero answer.  Therefore, for arbitrary $\vec k_i$, we do
not expect the theory to be at a zero of the $\beta$-function and we would
expect the coupling constant of $e^{i\vec k\cdot\vp(t)}$ to flow.  In fact,
in reference \klebanov, the $\beta$-function for arbitrary tachyon
potentials and zero gauge-field was calculated, and the potential which
satisfies the $\beta$-function at one-loop does not satisfy it at higher
orders \Larsnote.
This calculation, though, does not apply to the special case of the
cosine potential given in equation \DHMVdef, because the integrals
with the regulator in reference \klebanov\ are undefined in
the limit as
$\aab k_\mu^2\to 1$ for any individual {\it component} of each $\vec k$.
For the special
choice of potential given in equation \DHMVdef, whenever $\aab k_x^2 =
\aab k_y^2 = 1$ and $\beta/\alpha \in Z$, we can say much more if we use the
regulator defined in this paper instead.

According to references \cgcdef\ and \cff, we know that
in this case the free energy has no logarithmic
divergences.  In reference \freed, we prove that the free energy of
charge zero graphs goes as $1/\epsilon$ (with no logarithmic subdivergences.)
This means that all our approximations for charge-0 graphs are valid, and
they satisfy the Ward identity.
Similar calculations \charged\ prove that for all charged graphs, the free
energy, $F^*$, is always finite, as we expected, and that all the
approximations we made in calculating the flow for the coupling constant of
$e^{\vec k\cdot\vp(t)}$ are valid.  Thus $\epsilon \partial
F^*/\partial\epsilon =0$, as $\epsilon$ goes to zero, for all the charged
graphs.  The only difference between the graphs for $F^*$ and $\Gamma$ is
that in the latter we must subtract the 1PR graphs.  For the critical cosine
potential, using the results of reference \freed, it is straightforward
to show that all such 1PR graphs give $0$ contribution to $\Gamma$ and the
Ward identity as $\epsilon \to 0$.  We conclude that this potential is at a
zero of the $\beta$-function and satisfies the Ward identity to {\it all}
orders in $\alpha'$, at every order in $V$.

\newsec{Conclusions}
In this paper, we have shown that when the non-local, 1-D field theory has
a constant gauge field and arbitrary scalar potential, the reparametrization
invariance Ward identities are equivalent to the $\beta$-function
as the cutoff goes to zero.  One consequence of this result is that if the
theory is scale invariant, it also exhibits an infinite set of symmetries
which come from the reparametrization invariance of the underlying string
theory.  The proof presented in this paper is rather involved, so we hope
in the future one might find a simpler way to demonstrate the equivalence.
Also, the question still remains of whether, for {\it any} boundary state or
dissipative quantum system, scale invariance always implies that the Ward
identities are satisfied.

Secondly, combined with results from reference \freed\ and
generalizations, we have proved that the cosine potential of equation
\DHMVdef\ with $\aab {k_x}^2 = \aab {k_y}^2 =1$ and
$\beta/\alpha \in Z$ satisfies
the Ward identity.  Therefore, for this particular choice of on-shell
tachyon potential, the one-loop solution to the $\beta$-function remains a
solution to all orders in $V_0$ and $\alpha'$.  This differs from the case
for the general on-shell open string tachyon potential (as in references
\klebanov\
and \Larsnote) and for the sine-Gordon theory, where the $\beta$-function
gets corrections at higher orders in $V_0$, so that the value of the period
flows.

These results demonstrate that at these critical values of the cosine
potential and gauge field, the system is a conformal field theory, and, if
we include the remaining 24 dimensions, should give solutions to open string
theory with non-trivial backgrounds.  These backgrounds are of interest,
firstly because, using the methods in \freed, we can find exact solutions
for some of the correlation functions and can easily do computations to low
orders in $V_0$.  Secondly, there is a whole fractal network of
circles in the $\alpha-\beta$ plane where one-loop renormalization group
calculations and duality symmetries show that dissipative Hofstadter
model is critical.  We still do not know if these theories are really
critical to all orders in the potential.  If they are, the results in
this paper would suggest that these theories also give solutions to string
theory. Thus we would have a complex picture of what happens in open string
theory as we vary some of the background fields.

To apply the results in this paper to dissipative quantum systems, we must
integrate over the zero mode of $\vec X$ in equation \Wdef.  Then the Ward
identities for $n=0,\pm1$ says that the correlation functions for the
special critical theories are SU(1,1) covariant in addition to
satisfing the remaining Ward identities.  This means that the critical
theory is not only invariant under scaling and time translations, but also
under taking the time, $t$, to $1/t$.  From the point of view of
dissipative quantum systems, this enhanced symmetry is unexpected.  These
results are helpful in solving for the correlation functions of the
multi-critical dissipative Hofstadter model, but, by themselves, do not
contain enough information for solving the theory.   In a future
paper, we will combine these results with other symmetries of the dissipative
Hofstadter model to find exact solutions for some of the correlation
functions.

\newsec{Acknowledgements}
I would like to thank C. Callan, J. Cohn, A. Felce and S. Axelrod
for many helpful discussions.

\listrefs
\end